\documentclass{jfm}
\usepackage{hyperref} \hypersetup{colorlinks=true, allcolors=blue}
\usepackage{mathtools}
\usepackage[ruled,vlined,lined,linesnumbered]{algorithm2e}
\usepackage{stackrel}
\usepackage{multirow,bigdelim}
\usepackage{adjustbox}
\usepackage[dvipsnames]{xcolor}
\usepackage{tikz}
\usetikzlibrary{mindmap,shadows}

\newcommand\ie{{\textit{i.e.}}}
\newcommand\RR{\color{Black}}

\title{Image-based flow decomposition using empirical wavelet transform}
\shorttitle{Image-based flow decomposition using EWT}

\author{Jie Ren\aff{1}
		, 
            Xuerui Mao\aff{1}\corresp{\email{xuerui.mao@nottingham.ac.uk}}
            \and
            Song Fu\aff{2}}
\shortauthor{J. Ren, X. Mao and S. Fu}
\affiliation{\aff{1}Department of Mechanical Engineering, Faculty of Engineering, University of Nottingham, Nottingham, NG7 2RD, UK
		\aff{2}School of Aerospace Engineering, Tsinghua University, Beijing, 100084, China}

\graphicspath{{Figures/}}
\begin{document}
\maketitle

\begin{abstract}
We propose an image-based flow decomposition developed from the two-dimensional (2D) tensor empirical wavelet transform (EWT) \citep{Gilles2013}. The idea is to decompose the instantaneous flow {\RR data, or its visualisation,} adaptively according to the averaged Fourier supports for the identification of spatially localised structures. The resulting EWT modes stand for the decomposed flows, and each accounts for part of the spectrum, illustrating fluid physics with different scales superimposed in the {\RR original flow}. {\RR With the proposed method, decomposition of an instantaneous 3D flow becomes feasible without resorting to its time series. Examples first focus on the interaction between a jet plume and 2D wake, where only experimental visualisations are available. The proposed method is capable of separating the jet/wake flows and their instabilities. Then the decomposition is applied to an early stage boundary layer transition, where direct numerical simulations provided a full data-set. The tested inputs are the 3D flow data and its visualisation using  streamwise velocity \& $\lambda_{2}$ vortex identification criterion. With both types of inputs, EWT modes robustly extract the streamwise-elongated streaks, multiple secondary instabilities and helical vortex filaments. Results from bi-global stability analysis justify the EWT modes that represent the streak instabilities. In contrast to Proper Orthogonal Decomposition or Dynamic Modal Decomposition that extract spatial modes according to energy or frequency, EWT provides a new strategy as to decompose an instantaneous flow from its spatial scales. }

\end{abstract}


\section{Introduction}\label{S1}
The age of big data witnesses an expeditious generation and accumulation of flow data, both numerically and experimentally \citep{Duraisamy2019}. Even standing on top of data, on most occasions, it remains not a simple task to understand a fluid flow due to the broad range that the flow may comprise in its temporal and spatial scales. {\RR On the other hand,  complete data-set may not always be available. The extraction of concerned knowledge from projected flow data (\eg from flow visualisations) represents immense potentials for practical applications \citep{raissi2020hidden}.}

\begin{figure}
\begin{center}
\begin{tikzpicture}
\path[small mindmap,concept color=black!40,text=white,
    	every node/.style={concept, circular drop shadow},
    				root/.style      = {concept color=Black!60, font=\bfseries, text width=2cm},
    extra concept/.append style      = {concept color=Fuchsia,text width=1.4cm},
    	level 1 concept/.append style= {sibling angle=180,level distance=3cm,inner sep=0pt},
    	level 2 concept/.append style= {level distance=2cm}]
node[root] {Data-based modal analysis} [clockwise from=0]
    child[concept color=OliveGreen] 
    	{
		node (POD){POD} [sibling angle=30,clockwise from=60]
		child { node (SPOD) {spectral POD }}
		child { node(BPOD){balanced POD}}
		child { node{......}}
    	}
    child[concept color=NavyBlue] 
    	{
		node[concept] (DMD){DMD}[sibling angle=30,counterclockwise from=120]
		child { node[concept] (RDMD){recursive DMD}}
		child { node[concept] (spDMD) {sparsity promoting DMD}}
		child { node[concept] {......}}
    	}
node[extra concept] (SB)at (-2.0,2.2){Stability analysis}
node[extra concept] (RS)at (0.4,2.3){Resolvent analysis}
node[extra concept] (KPM)at (-1,-2){Koopman analysis};
\draw [concept connection]  (DMD) edge (KPM)
						(DMD) edge (RS)
						(DMD) edge (SB)
						(RS) edge (SB)
						(RS) edge (SPOD);
\end{tikzpicture}
\end{center}
\caption{A sketch of the state-of-art data-based modal analysis techniques for fluid flows.}
\label{Mind}
\end{figure}
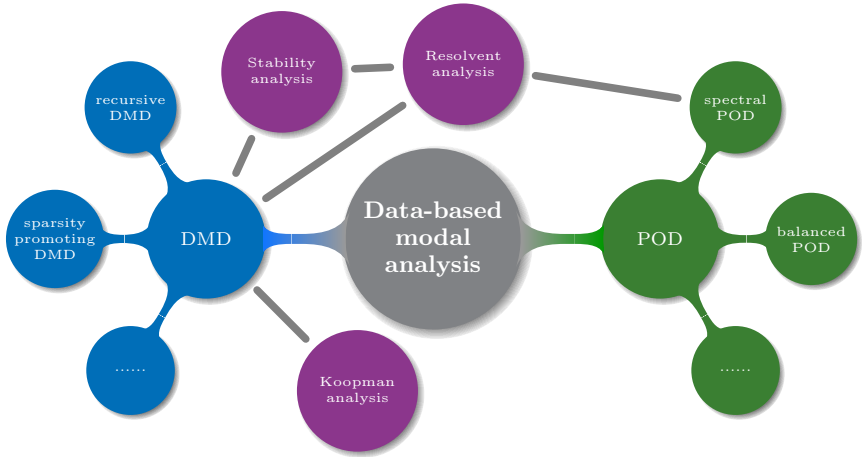

The development of model reduction and modal analysis techniques has played an essential role to promote this effort \citep[see reviews by][]{rowley2017model,Taira2017,Taira2019}. Figure~\ref{Mind} provides a general view of the well-developed data-based modal analysis techniques. With these approaches, a flow field has been decomposed into modes ranked by their inherent properties, \eg energy, frequency or growth rate. For example, Proper orthogonal decomposition (POD) \citep{lumley1967} determines the optimal set of modes that represent most of the energy based on the $L_2$ norm. The optimality lies both in the least possible number of modes to represent the signal and in the minimisation of error between the signal and its truncated representation.  
Dynamic mode decomposition (DMD) \citep{Schmid2010}, on the other hand, arranges modes in the order of their dynamical importance that is measured by a characteristic frequency and growth rate. 
In essence, DMD is a finite-dimensional approximation to the Koopman operator \citep{Rowley2009}. For nonlinear problems, choosing a suitable set of observables as input is critical to maintain the link to the Koopman operator.  
Various improvements to the standard POD and DMD have been proposed. For example, balanced POD \citep{rowley2005model} weighs  controllability and observability of a state by forming a biorthogonal set, and therefore, is more suitable for non-normal systems. Spectral POD \citep{towne2018spectral} reconsiders POD in the frequency domain, providing orthogonal modes (still ranked by energy) at discrete frequencies. 
Sparsity promoting DMD \citep{jovanovic2014sparsity} reduces in the dimensionality of the full rank decomposition, and Recursive DMD \citep{noack2016recursive} achieves orthogonality, an essential property of POD.
Both POD and DMD are purely data-driven, in contrast to the model-based linear global stability analysis \citep{Theofilis2011} and resolvent  analysis \citep{Mckeon2010}, which require an accurate base flow as well as the construction of the linearised operator.
From the view point of the stochastic process, \citet{towne2018spectral} showed the inherent connection between DMD, resolvent analysis and Spectral POD, indicating their common ground in mathematics.

The above reviewed data-based decompositions aim at extracting dominant structures from a series of snapshots, while in the present work, an image-based technique will be introduced to obtain meaningful structures from instantaneous flow data, even though it can also apply to a sequence of snapshots. This technique is cost-effective and supports real-time analysis of flows upon observations. 


The image-based flow decomposition hinges on the basis of the empirical wavelet transform (EWT) \citep{Gilles2013,Gilles2020}. Wavelet transform was recognised as a powerful technique in fluid mechanics not long after its birth \citep[see early review by][]{farge1992wavelet}. The pioneering work of \citet{Meneveau1991} brings turbulence to the orthonormal wavelet space. Cognition of turbulent kinetic energy, energy transfer and nonlinear interactions was obtained by physically interpreting the wavelet coefficients. In the examination of fluid flow data, wavelet analysis has been widely used as a flexible ``microscope'' discriminating scales and positions simultaneously. 
{\RR The coherent vortex extraction (CVE) proposed by \citet{farge2001coherent} splits turbulent flows into coherent and incoherent parts. CVE achieves its goal by projecting the vorticity field onto an orthogonal wavelet basis, and a threshold on the resulting wavelet coefficients is then specified to identify the coherent component.}
In the last decade, wavelet-based numerical algorithms and turbulence modelling have been developed \citep{Schneider2010}. Their strengths lie in the competence to unambiguously identify and isolate localised, dynamically dominant flow structures which improve the efficiency of the computation. EWT provides the first mathematically rigorous method to adaptively analyse a signal \citep{Gilles2013}. The principal idea is to build a set of adaptive filter banks based on the property of the input signal itself, such that meaningful modes are extracted, each possessing a particular section of the Fourier supports. This feature matches well with the requirement to extract modes from instantaneous flow data or its visualisation.

{\RR In \S \ref{S2}, the formulation and algorithm of EWT for flow decompositions are introduced.  \S \ref{S3} provides two examples as EWT being applied to 3D instantaneous flow data and its visualisation, followed by the comparison with the other methods in \S\ref{S41}. The study is concluded in \S \ref{S4}.}


\section{Imaged-based flow decomposition using EWT}\label{S2}
{\RR We term the complete flow data-set or the visualisation (image or video) as a function $f(x,y,z;t)$. Here $x$, $y$, and $z$ represent the spatial coordinates, and the optional $t$ denotes time. Note that flow visualisations project data onto a 2D image or video: $f(x,y;t)$.} The visualisation may be qualitative and aided with auxiliary materials (\eg smoke, sensitive coatings), be quantitatively made with state-of-art facilities (\eg particle image velocimetry), or from numerical simulations (\eg using $\lambda_2$ criterion \citep{Jeong1995} to identify vortex structures). We decompose the flow using two-dimensional (2D) tensor EWT \citep{Gilles2014} as summarised in Algorithm \ref{A1}. 

EWT defines its scaling function $\hat{\phi}_{m}$ and empirical wavelets $\hat{\psi}_{m}$ (see Appendix \ref{appA}) in a normalised Fourier axis with $2\pi$ periodicity. The set $\{ \phi_{1},\left\{ \psi_{m}\right\} _{m=1}^{M-1}\}$ is a tight and orthogonal frame of $L^{2}\left(\mathbb{R}\right)$. Consequently, the ‘energy’ of the signal is conserved by the extracted EWT modes. The empirical wavelet transform of a signal $f(t)$ is obtained by
\begin{equation}
\left.\begin{array}{c}
\mathcal{W}_{1}=\left\langle f,\phi_{1}\right\rangle =\int f\left(t\right)\phi_{1}\left(\tau-t\right)\mathrm{d}\tau=\left(\hat{f}\left(\omega\right)\hat{\phi}_{1}^{\dagger}\left(\omega\right)\right)^{\vee}\\
\mathcal{W}_{m+1}=\left\langle f,\psi_{m}\right\rangle =\int f\left(t\right)\psi_{m}\left(\tau-t\right)\mathrm{d}\tau=\left(\hat{f}\left(\omega\right)\hat{\psi}_{m}^{\dagger}\left(\omega\right)\right)^{\vee},~m=1,...M-1
\end{array}\right\}, 
\end{equation}
where $\left\langle \right\rangle$ stands for the inner product, $\dagger$ is the complex conjugate, $\wedge$ and $\vee$ indicate the Fourier transform and its inverse. 

\begin{figure}
\begin{center}
\includegraphics[width=0.8\linewidth]{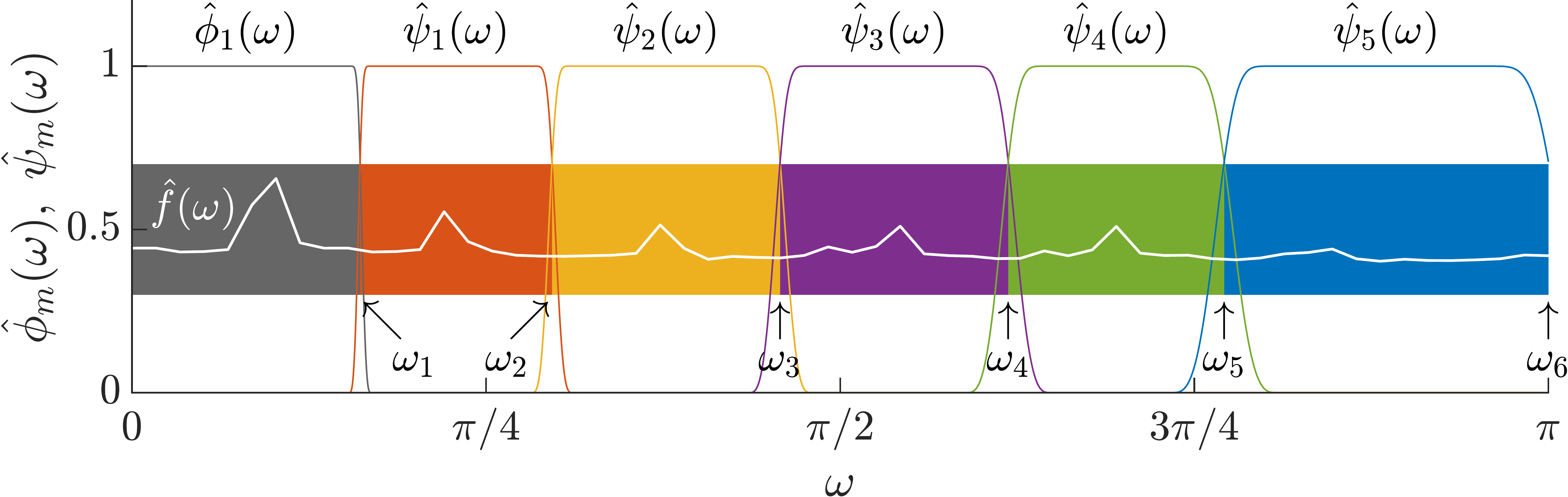}
\put(-64,52){\rotatebox{0}{\textcolor{white}{skeleton mode}}}
\put(-274,41){\rotatebox{0}{\textcolor{white}{shadow}}}
\put(-271,33){\rotatebox{0}{\textcolor{gray}{mode}}}
\end{center}
\caption{A sketch in the Fourier space of the signal $\hat{f}(\omega)$, the detected supports $\{\omega_m\}$, and the filter bank $\{\hat{\phi}_1(\omega)$, $\hat{\psi}_m(\omega)\}$. $\omega\in\left[0,\pi\right]$ and abides by the Nyquist–Shannon sampling theorem. The Fourier supports are determined from the spectrum of the signal $\hat{f}(\omega)$ such that each EWT mode amounts to part of the spectrum. The first and the last EWT modes, $\mathcal{W}_1$ and $\mathcal{W}_{M}$, are termed the shadow mode and the skeleton mode, respectively.}
\label{F1_sketch}
\end{figure}

\SetKwInput{KwInput}{Input} 
\SetKwInput{KwOutput}{Output} 
\begin{algorithm}\label{A1}
\DontPrintSemicolon
\caption{\RR Flow decomposition using 2D tensor empirical wavelet transform}
\SetAlgoLined
\KwInput{$f(x,y,z;t)$ with $N_x$, $N_y$, $N_z$ and $N_t$ points in corresponding indices. If $t$ (optional) is present, the input is a 3D unsteady flow or its visualisation (video); $M_x$, $M_y$, the number of desired filters/modes in the direction of $x$ and $y$. Note that we show the decomposition algorithm in the $x$ and $y$ directions, while in practical applications, the two directions can be chosen from $x$, $y$ and $z$.}
For inputs with flow visualisations, detect the principle direction of $f(x,y,z;t)$(details can be found in \S\ref{Ex1}). Rotate $f$ such that the principle direction is in line with $x$ or $y$;\;
In the $x$ direction, compute its FFT to obtain the $x$-averaged spectrum. For the input with $N_t$ snapshots, the spectrum is also averaged in time:
$$\hat{f}_{x}\left(\omega_{x}\right)=\frac{1}{N_yN_zN_{t}}\stackrel[j=1]{N_{y}}{\sum}\stackrel[k=1]{N_{z}}{\sum}\stackrel[t=1]{N_{t}}{\sum}\left|\hat{f}\left(j,k,\omega_{x};t\right)\right|;$$\;\vspace{-5mm}
Similarly, obtain the $y$-averaged spectrum: 
$$\hat{f}_{y}\left(\omega_{y}\right)=\frac{1}{N_{x}N_{z}N_{t}}\stackrel[i=1]{N_{x}}{\sum}\stackrel[k=1]{N_{z}}{\sum}\stackrel[t=1]{N_{t}}{\sum}\left|\hat{f}\left(i,k,\omega_{y};t\right)\right|;$$\;\vspace{-5mm}
Perform the Fourier boundary detection on $\hat{f}_{x}\left(\omega_{x}\right)$ and $\hat{f}_{y}\left(\omega_{y}\right)$ respectively;\;
Build filter banks $\mathcal{B}_x=\{ \phi_{1}^x,\left\{ \psi_{m}^x\right\} _{m=1}^{M_x-1}\}$ and $\mathcal{B}_y=\{ \phi_{1}^y,\left\{ \psi_{m}^y\right\} _{m=1}^{M_y-1}\}$ using \eqref{ewt1} and \eqref{ewt2};\;
Filter $f$ along the $x$ direction with  $\mathcal{B}_x$, resulting in $M_x$ outputs;\;
Filter each output in step 6 along $y$ direction using $\mathcal{B}_y$;\;
If necessary, reconstruction of the input $f(x,y,z;t)$ is performed by inverting the $y$ and $x$ filtering successively following 
$$f=\left(\mathcal{\hat{W}}_{1}\hat{\phi}_{1}+\stackrel[m=1]{M-1}{\sum}\mathcal{\hat{W}}_{m+1}\hat{\psi}_{m}\right)^{\vee}.
$$\;\vspace{-5mm}
\KwOutput{EWT modes: $\mathcal{W}_{i,j}(x,y,z;t)$ with $i = 1,...,M_{x}$, $j = 1,...,M_{y}$ .}
\end{algorithm}

We show in figure~\ref{F1_sketch} an example of the adaptive filter bank. The signal in Fourier space  $\hat{f}(\omega)$ is shown with a white line. In this study, the Fourier supports $\{\omega_m\}$ are adaptively determined such that local maxima of $\hat{f}(\omega)$ are retained in different Fourier segments (indicated with different colours).  As a result, the extracted modes strategically capture components of the input signal, each standing for a section of the spectrum. {\RR The idea of 2D tensor EWT is to perform EWT successively in two directions, with the corresponding filter banks built according to the spatial-averaged spectrum.} When flow visualisation is used as an input, it is essential to have the principle direction of the flow (\eg direction of the mean flow) coincide with either direction of decomposition, such that the averaged spectrum maximises its physical relevance. This corresponds to step 1 of algorithm \ref{A1} and will be detailed with an example in \S \ref{Ex1}.

\section{Applications and results}\label{S3}
We show two applications of flow decomposition using EWT: the interaction between a 2D wake and a jet plume, and the early-stage boundary layer transition subject to free-stream turbulence. The two examples serve to demonstrate the applicability {\RR both to flow data and its visualisation}, irrespective of their experimental or numerical origin.

\subsection{Interaction between a jet plume and a 2D wake (images from experiments)}\label{Ex1}
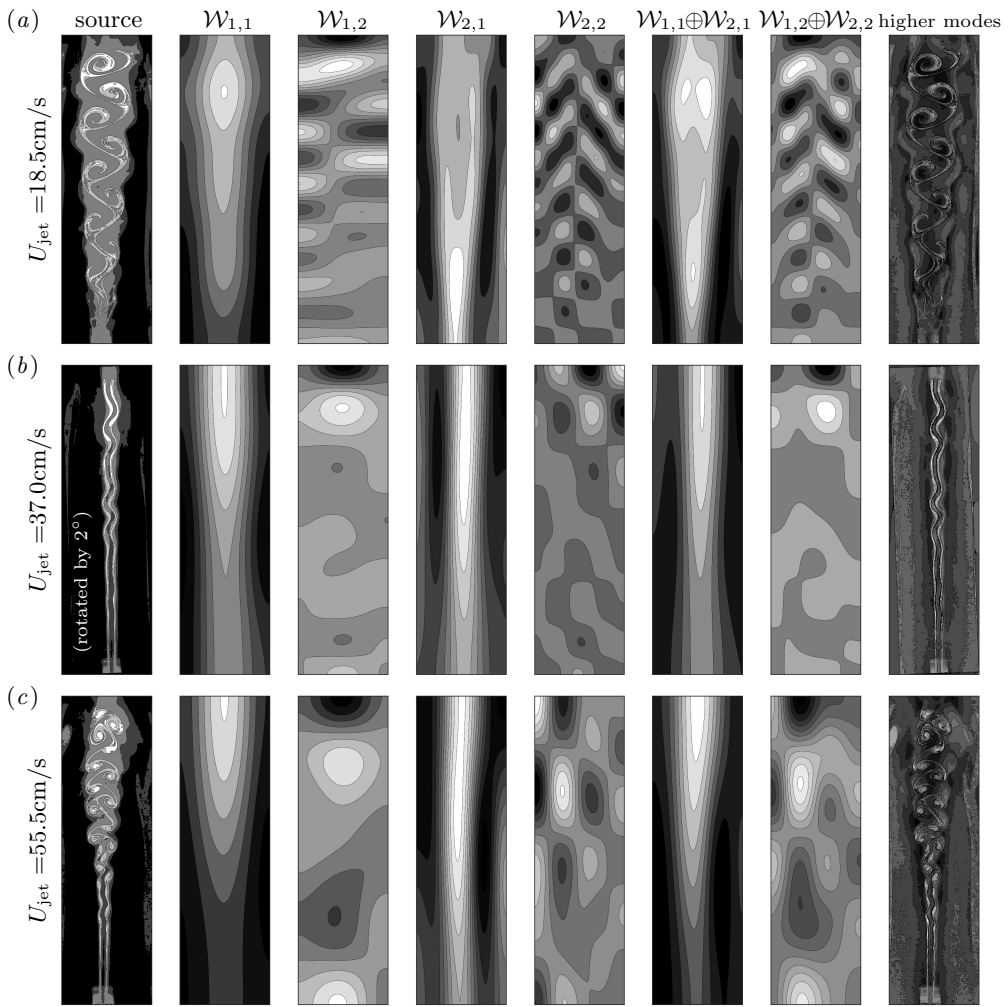
\begin{figure}
\begin{center}
\includegraphics[trim={70mm 23mm 70mm 0},clip,width=0.95\linewidth]{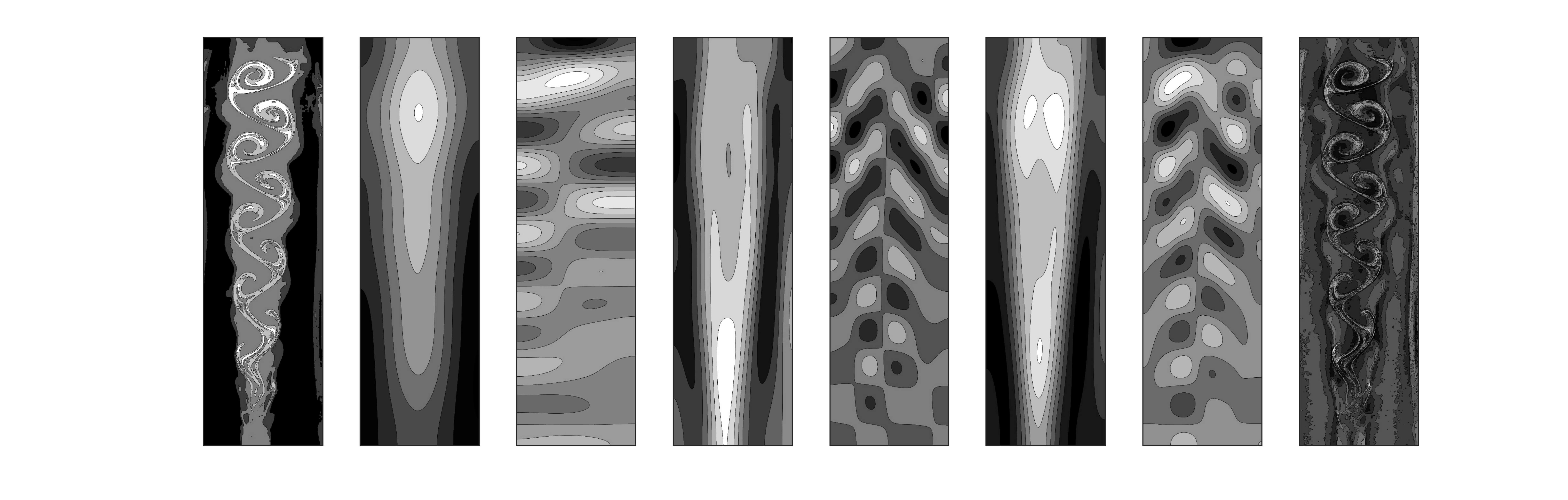}
\put(-368,121){({\it a\hspace{1pt}})}
\put(-342,121){source}
\put(-294,121){$\mathcal{W}_{1,1}$}
\put(-251,121){$\mathcal{W}_{1,2}$}
\put(-206,121){$\mathcal{W}_{2,1}$}
\put(-161,121){$\mathcal{W}_{2,2}$}
\put(-131,121){$\mathcal{W}_{1,1}\hspace{-3pt}\oplus\hspace{-3pt}\mathcal{W}_{2,1}$}
\put(-085,121){$\mathcal{W}_{1,2}\hspace{-3pt}\oplus\hspace{-3pt}\mathcal{W}_{2,2}$}
\put(-040,121){\scriptsize higher modes}
\put(-360,33){\rotatebox{90}{$U_{\rm jet}=$18.5cm/s}}
\\
\includegraphics[trim={70mm 23mm 70mm 10mm},clip,width=0.95\linewidth]{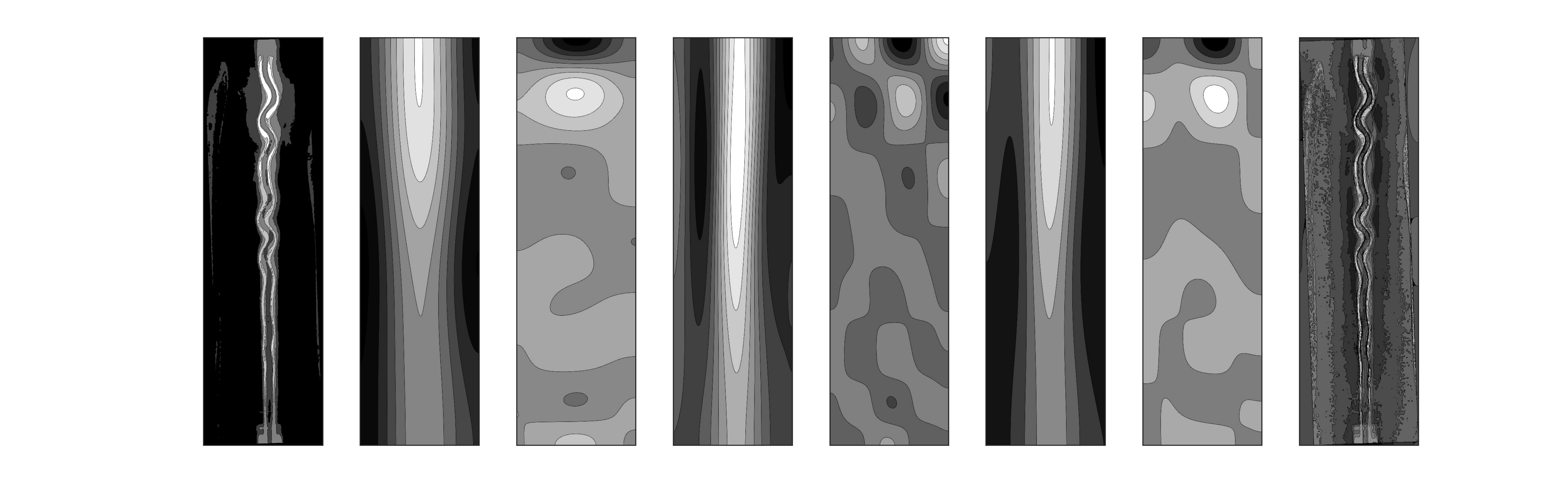}
\put(-368,115){({\it b\hspace{1pt}})}
\put(-343,10){\rotatebox{90}{\textcolor{white}{\scriptsize (rotated by $2^{\circ}$)}}}
\put(-360,33){\rotatebox{90}{$U_{\rm jet}=$37.0cm/s}}
\\
\includegraphics[trim={70mm 23mm 70mm 10mm},clip,width=0.95\linewidth]{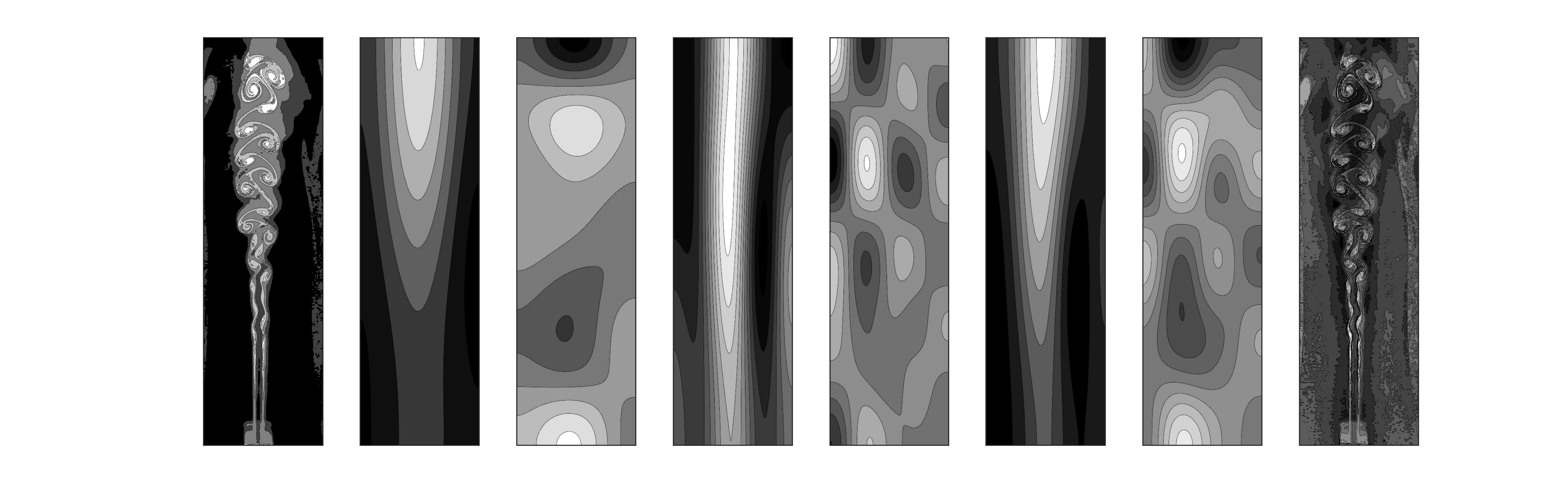}
\put(-368,115){({\it c\hspace{1pt}})}
\put(-360,33){\rotatebox{90}{$U_{\rm jet}=$55.5cm/s}}
\end{center}
\caption{Image-based flow decomposition of the interaction between a jet plume and a 2D wake.  (a) $U_\mathrm{jet}=18.5$cm/s; (b) $U_\mathrm{jet}=37.0$cm/s with the source image rotated $2^{\circ}$ counterclockwise; (c) $U_\mathrm{jet}=55.5$cm/s. The wake is subject to an annular air flow at a constant velocity of 27.5 cm/s.}
\label{F5_jet2}
\end{figure}

In this example, the decomposition is applied to the experimental observation of the interaction between a jet plume and a 2D wake \citep{Roquemore1988}. The slot jet is located in the centre of the rectangular face of a bluff body. The flow is visualised with reactive Mie scattering technique, where streamlines are highlighted by TiO$_2$ particles that are spontaneously formed by the reaction of TiCl$_4$ in the jet with water in the wake. The flow visualisation is shown in the first column of figure~\ref{F5_jet2} with jet velocity increasing from 18.5 cm/s in panel (a) to 55.5 cm/s in (c).  As introduced in \citet{Roquemore1988}, the flow is initially dominated by the wake flow when the shear layer velocity of the jet $U_\mathrm{jet}$ is smaller. Along with the increase of $U_\mathrm{jet}$ in panel (b), wake instabilities are considerably prohibited, and only wavy structures are observed. In panel (c), the flow is dominated by the jet instability as indicated by the change in the direction of rotation of the vortices. 

The flow decomposition is performed with $M_x=M_y=3$, leading to nine EWT modes, ${\mathcal{W}_{i,j}}$ ($i,j=1,2,3$). The lower modes (columns 2 $\sim$ 5 of figure~\ref{F5_jet2}) well isolate key components of the flows. In particular, mode $\mathcal{W}_{1,1}$ (the shadow mode), comprising the lower-end spectrum both in the flow and traverse directions, can be used to detect the principal direction of the flow.  Mode $\mathcal{W}_{2,1}$ possesses mid-spectrum in the traverse direction and features the mean flow. Representing the mid-spectrum of the flow direction, mode $\mathcal{W}_{1,2}$ and $\mathcal{W}_{2,2}$ present the flow instabilities with  $\mathcal{W}_{2,2}$ accounting for higher wavenumbers in the traverse direction. Comparing the three cases adheres to the intention of the experiment: an increase of the jet velocity leads to the suppression of wake instabilities (panel b), and promotion of the jet instability downstream (panel c), as it is supported by $\mathcal{W}_{2,2}$ in the three panels.  It is also helpful to inspect the flow by combining certain modes: $\mathcal{W}_{1,1}\oplus\mathcal{W}_{2,1}$ and $\mathcal{W}_{1,2}\oplus\mathcal{W}_{2,2}$.  The combination is obtained by performing an inverse EWT with corresponding modes. As it is shown, $\mathcal{W}_{1,1}\oplus\mathcal{W}_{2,1}$ and $\mathcal{W}_{1,2}\oplus\mathcal{W}_{2,2}$ account for a wider spectrum in the traverse direction and differentiate in the flow direction, thus delivering a more comprehensive view of the mean flow and wake/jet instabilities respectively. Note that we have reconstructed the higher modes (the last column of figure~\ref{F5_jet2}) with $\mathcal{W}_{1,3}\oplus\mathcal{W}_{3,1}\oplus\mathcal{W}_{2,3}\oplus\mathcal{W}_{3,2}\oplus\mathcal{W}_{3,3}$. The higher modes amount for the rest of the spectrum and retain all the small scales of the visualisation. Note that by increasing the number of EWT modes, the higher modes (the skeleton) will contain less information of the input.

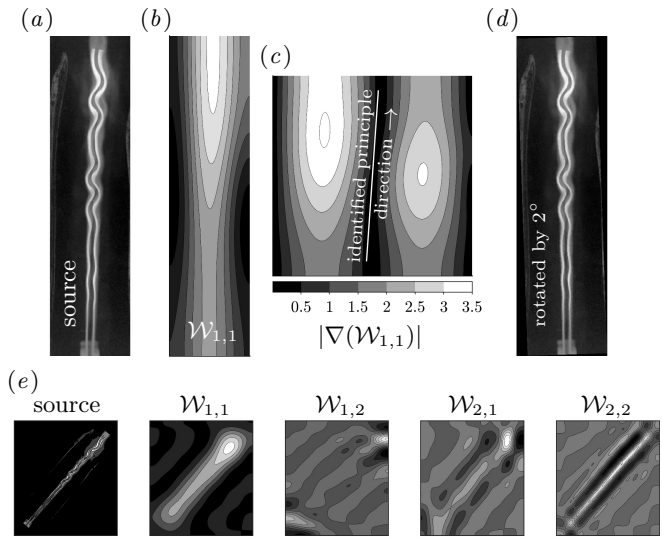
\begin{figure}
\begin{center}
\includegraphics[trim={0 15mm 0 0},clip,width=0.65\linewidth]{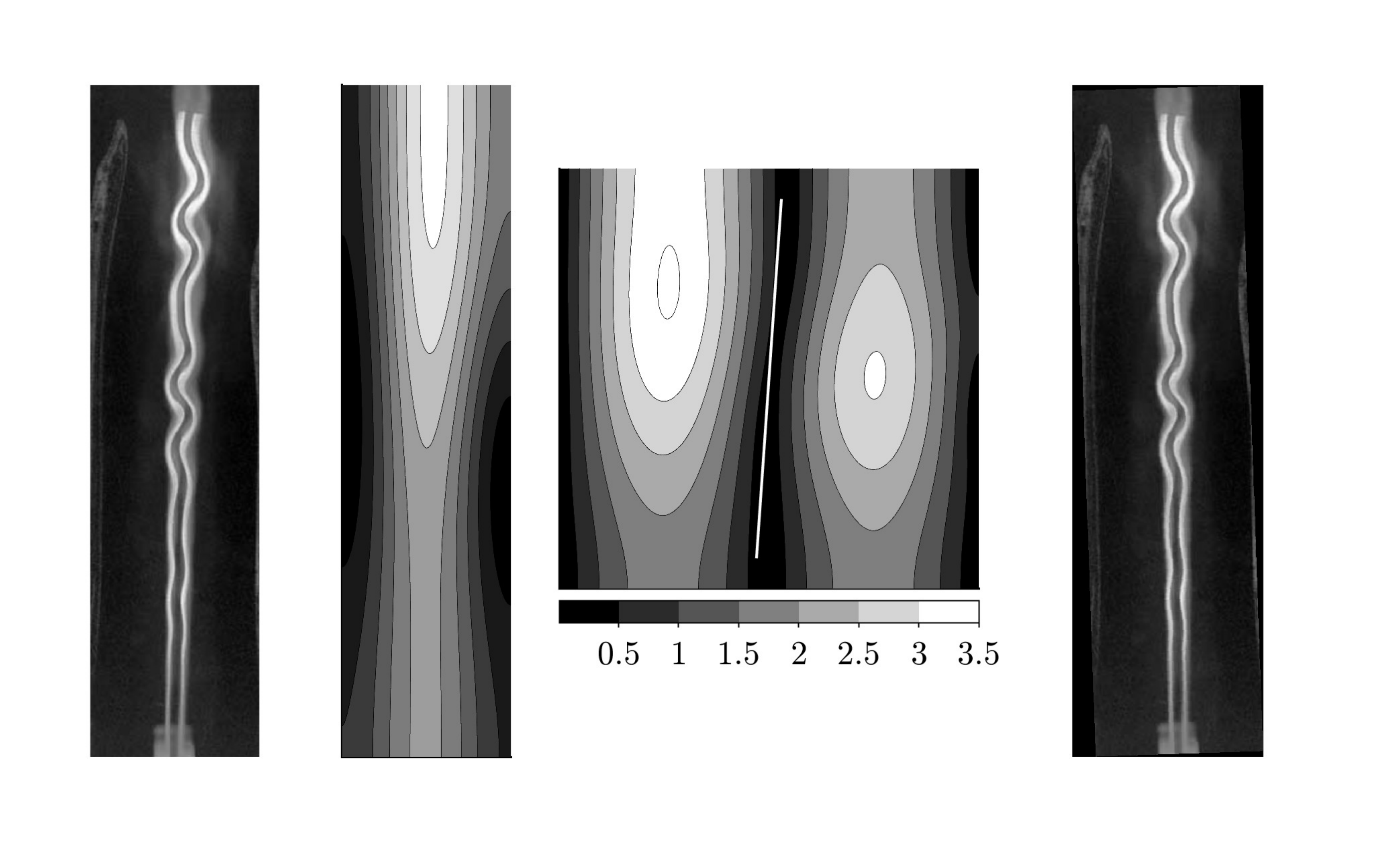}
\put(-245,128){({\it a\hspace{1pt}})}
\put(-200,128){({\it b\hspace{1pt}})}
\put(-155,114){({\it c\hspace{1pt}})}
\put(-70,128){({\it d\hspace{1pt}})}
\put(-228,15){\rotatebox{90}{\textcolor{white}{source}}}
\put(-53,15){\rotatebox{90}{\textcolor{white}{\scriptsize rotated by $2^{\circ}$}}}
\put(-182,11){\rotatebox{0}{\textcolor{white}{$\mathcal{W}_{1,1}$}}}
\put(-133,09){\rotatebox{0}{\textcolor{black}{$|\nabla(\mathcal{W}_{1,1})|$}}}
\put(-122,39){\rotatebox{86}{\textcolor{white}{\rm \scriptsize identified principle}}}
\put(-112,55){\rotatebox{86}{\textcolor{white}{\rm \scriptsize direction}}}
\put(-107,88){\rotatebox{86}{\textcolor{white}{$\rightarrow$}}}
\\
\includegraphics[trim={6cm 5cm 4cm 4cm},clip,width=0.65\linewidth]{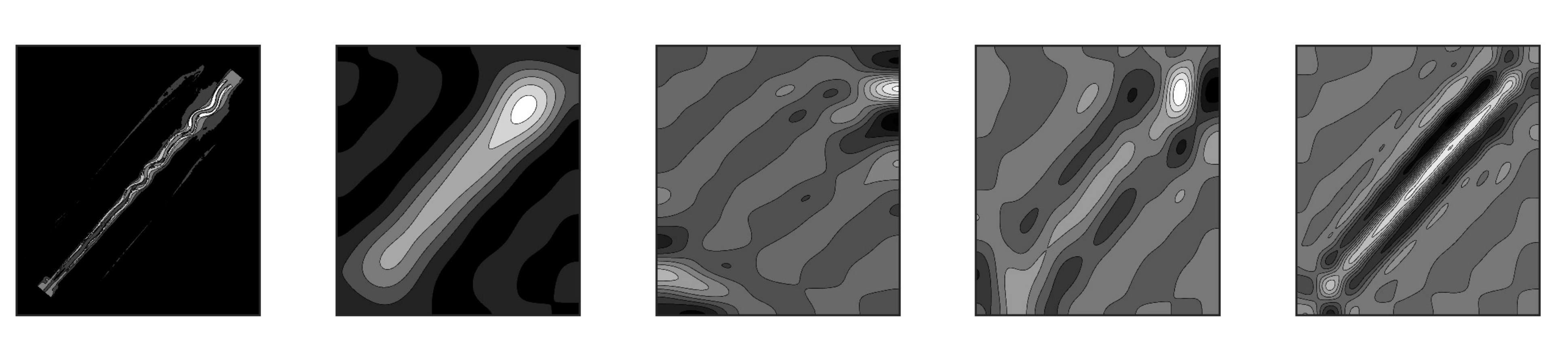}
\put(-250,60){({\it e\hspace{1pt}})}
\put(-240,51){source}
\put(-185,51){$\mathcal{W}_{1,1}$}
\put(-134,51){$\mathcal{W}_{1,2}$}
\put(-084,51){$\mathcal{W}_{2,1}$}
\put(-034,51){$\mathcal{W}_{2,2}$}
\end{center}
\caption{Identification of the principle direction. (a) The source taken from the $U_\mathrm{jet}=37.0$cm/s case of \citet{Roquemore1988}. (b) $\mathcal{W}_{1,1}$ of the original image. (c) Image gradient of $\mathcal{W}_{1,1}$. For better display of the identified principle direction, the $x$-$y$ aspect ratio is not to scale. (d) The rotated image for flow decomposition. (e) Flow decomposition based on the source with an inappropriate orientation.}
\label{F6_rotate}
\end{figure}

Recalling algorithm \ref{A1}, in 2D tensor EWT, the wavelet filter bank is built based on the averaged spectrum in the $x$ and $y$ directions of the signal. It is crucial that the flow's principal direction (for example, the direction of the mean flow, orientations of the geometry) coincides with these directions, such that the flow physics can be correctly separated.  In the 37.0cm/s case of the above visualisation, we have rotated the source counterclockwise by 2.0$^{\circ}$ before the EWT is applied (step 1 of Algorithm~\ref{A1}).  We take this case as an example to show how to accurately detect the principal direction. As shown in figure~\ref{F6_rotate}, the shadow mode $\mathcal{W}_{1,1}$, which is free from small scales, characterises the principle direction of the signal. This direction is thus numerically recovered by the gradient, $|\nabla(\mathcal{W}_{1,1}) |$. The principle direction is the direction along which the average gradient reaches a minimum, as shown in figure~\ref{F6_rotate}(c). Consequently, it is found that the source shall be rotated counterclockwise by 2.0$^{\circ}$. Figure~\ref{F6_rotate}(d) shows the rotated image as compared with figure~\ref{F6_rotate}(a). Panel (e) shows the decomposition when the source image has an inappropriate orientation at which $\mathcal{W}_{2,1}$ and $\mathcal{W}_{2,2}$ fail to capture the mean flow and instabilities.

\subsection{Early stage boundary layer transition (from direct numerical simulations)}\label{S32}

\begin{figure}
\begin{center}
\includegraphics[trim={0 1mm 4mm 0},clip,width=0.95\linewidth]{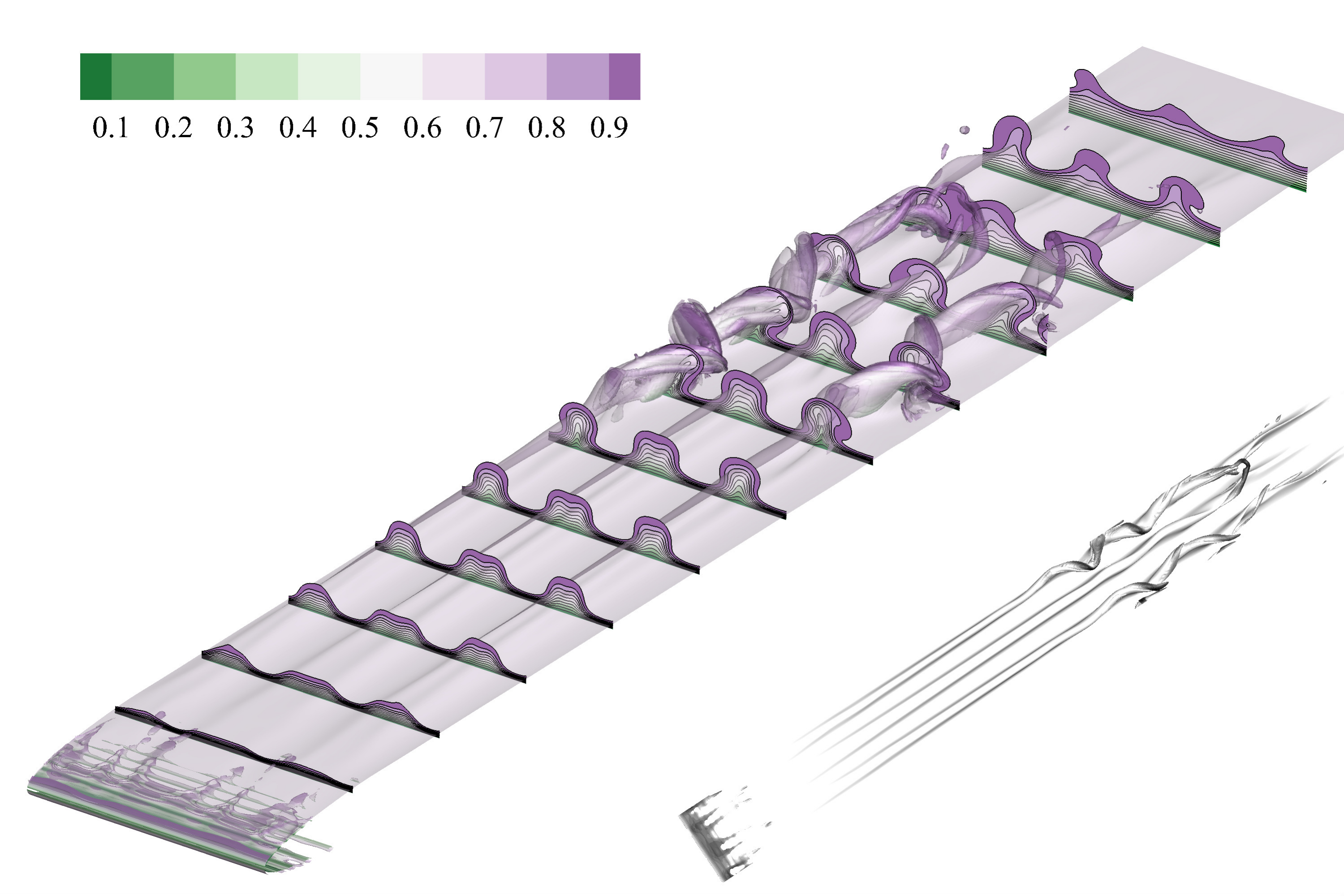}
\put(-345, 192){{Contour slices and iso-surface coloured}}
\put(-318, 182){{by streamwise velocity $u$}}
\put(-358, 40){\rotatebox{34}{\Large $\longrightarrow$}}
\put(-338, 55){\rotatebox{33}{freestream direction}}
\put(-113, 63){\rotatebox{33}{A}}
\put(-108, 55){\rotatebox{33}{B}}
\put(-103.5, 47){\rotatebox{33}{C}}
\put(-139, 34){\rotatebox{33}{streaks}}
\put(-116, 51){\rotatebox{33}{$\Big\{$}}
\put(-127, 90){\rotatebox{35}{\bf (3D Data)}}
\put(-95, 40){\rotatebox{35}{\bf (2D Image)}}
\put(-48, 130){\rotatebox{33}{\scriptsize hairpin}}
\put(-45, 124){\rotatebox{33}{\scriptsize vortices}}
\put(-32, 128){\rotatebox{-55}{\scriptsize$\rightarrow$}}
\put(-101,94){\rotatebox{-55}{\Large\textcolor{Orchid}{$\Rightarrow$}}}
\put(-280, 10){\rotatebox{33}{$x=10$}}
\put(-245, 33){\rotatebox{33}{$20$}}
\put(-223, 48){\rotatebox{33}{$30$}}
\put(-200, 62){\rotatebox{33}{$40$}}
\put(-175, 78){\rotatebox{33}{$50$}}
\put(-151, 93){\rotatebox{33}{$60$}}
\put(-127, 109){\rotatebox{33}{$70$}}
\put(-104, 124){\rotatebox{33}{$80$}}
\put(-80, 139){\rotatebox{33}{$90$}}
\put(-58, 153){\rotatebox{33}{$100$}}
\put(-34, 168){\rotatebox{33}{$110$}}
\put(-12, 182){\rotatebox{33}{$120$}}
\put(-354,16){\rotatebox{-17}{leading edge}}
\put(-200,30){\Large\textcolor{Orchid}{$\Rightarrow$}}
\put(-10,164){\rotatebox{-90}{\Large\textcolor{Orchid}{$\Rightarrow$}}}
\end{center}
\caption{A schematic diagram of imaged-based flow decomposition with the example of boundary layer transition. The instantaneous flow field (3D data) is visualised by iso-surfaces of $\lambda_2=-0.02$ and $u=0.8$ together with contours of $u=0.1,0.2,...,0.9$ in cross sections that are perpendicular to the freestream direction. Both the iso-surfaces and contours are coloured by $u$. We apply imaged-based flow decomposition on the grayscale 2D image that contains the same iso-surfaces. The image has a top view on which the flow data in the wall normal direction is projected. Within the computation domain, three discernible streaks are generated, and they become unstable downstream before giving rise to hairpin vortices.}
\label{F2_bypass1}
\end{figure}

Laminar-turbulent transition prompts a significant broadening of flow scales that ultimately takes the form of coherent structures. This example examines an incompressible boundary layer flow over a flat plate with an elliptic leading edge  (see figure~\ref{F2_bypass1}). To promote flow transition, free-stream turbulence (FST) from a nonlinear optimisation procedure (with the target that the maximum perturbation energy is reached at a designated time) is specified ahead of the leading edge. The width of the domain is chosen such that three discernible streamwise-elongated streaks are generated. A periodic boundary condition is employed in the spanwise direction. Further numerical details and set-up of the simulations can also be found in \citet{Bofu2019}. {\RR In this example, we first apply the algorithm to flow visualisation, followed by the decomposition with instantaneous 3D velocity data.}

The flow has been visualised through its pivotal structures, \ie, with iso-surfaces of $\lambda_2=-0.02$, $u=0.8$ and contour slices of  $u=0.1, 0.2 ,..., 0.9$ as presented in figure~\ref{F2_bypass1}, here $u$ stands for the streamwise velocity (normalised by the freestream value). A grey-scale top-view image with the same iso-surfaces serves as the input of the image-based flow decomposition. As can be seen, the 2D image features the generation of streaks and the formation of hairpin vortices.

\begin{figure}
\begin{center}
\hspace{4mm}
\begin{minipage}{0.48\textwidth}
\includegraphics[trim={0 17mm 0 0},clip,width=0.95\linewidth]{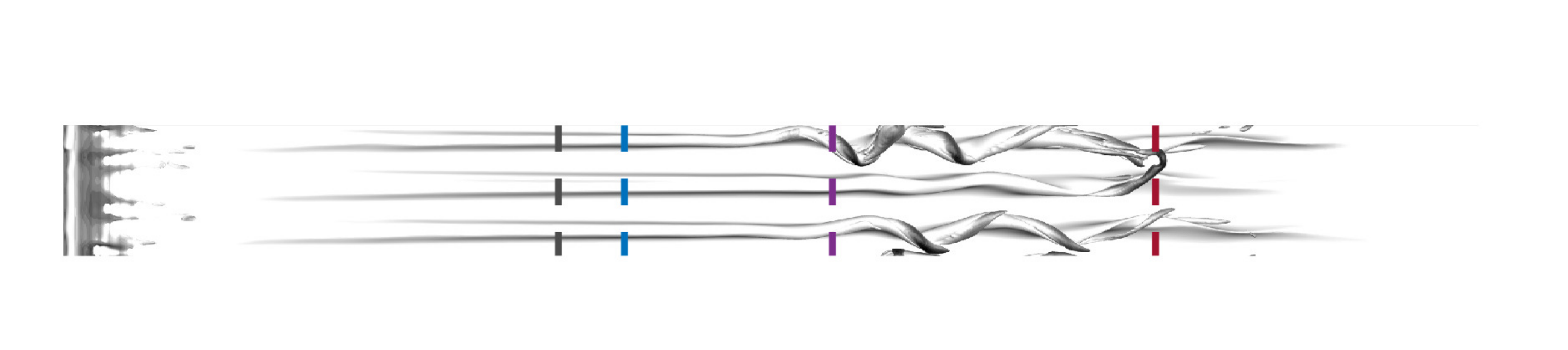}\\
\includegraphics[width=0.95\linewidth]{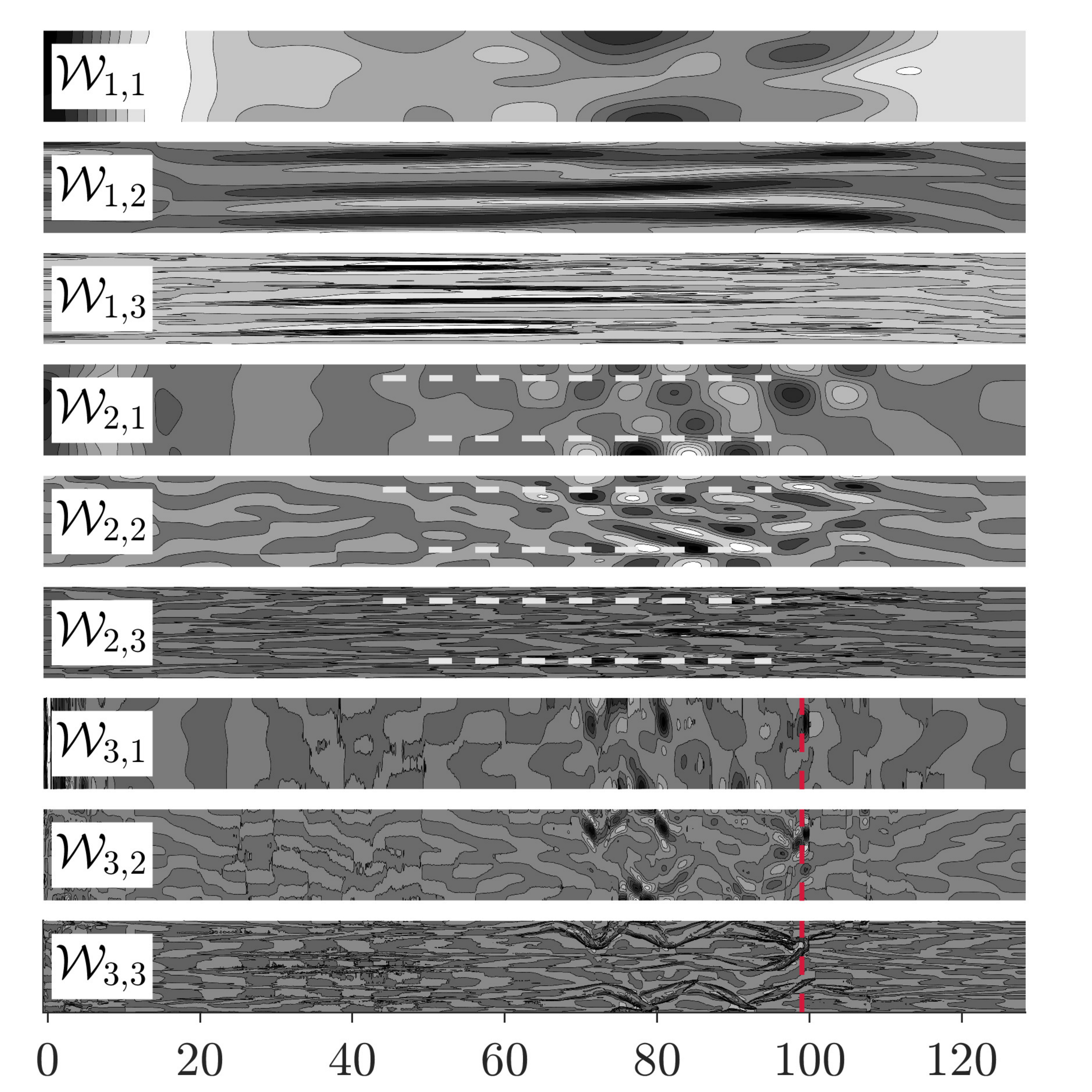}
\end{minipage}
\hspace{2mm}
\begin{minipage}{0.33\textwidth}
\includegraphics[width=1.0\linewidth]{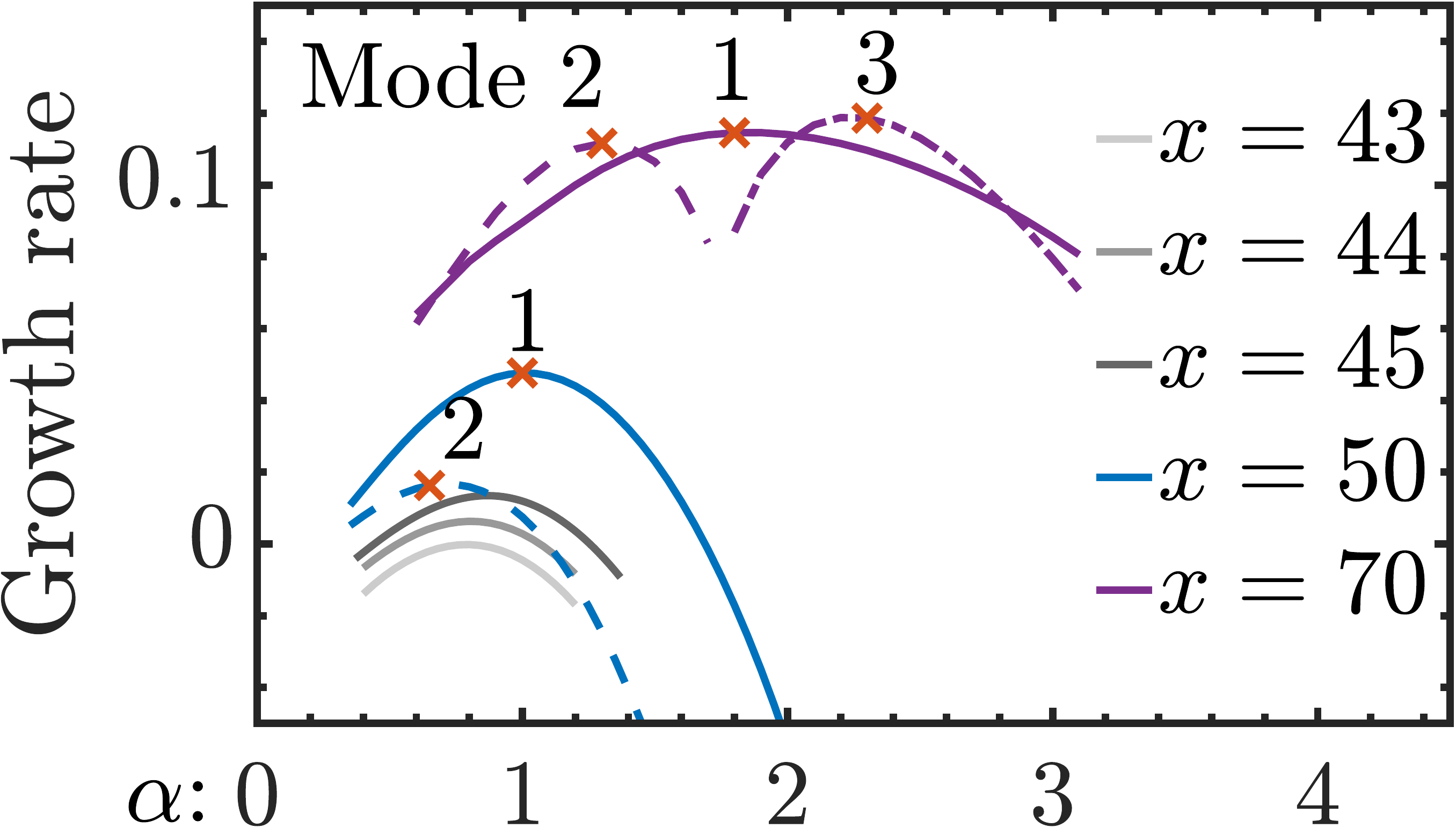}\vspace{10pt}\\
\includegraphics[width=1.0\linewidth]{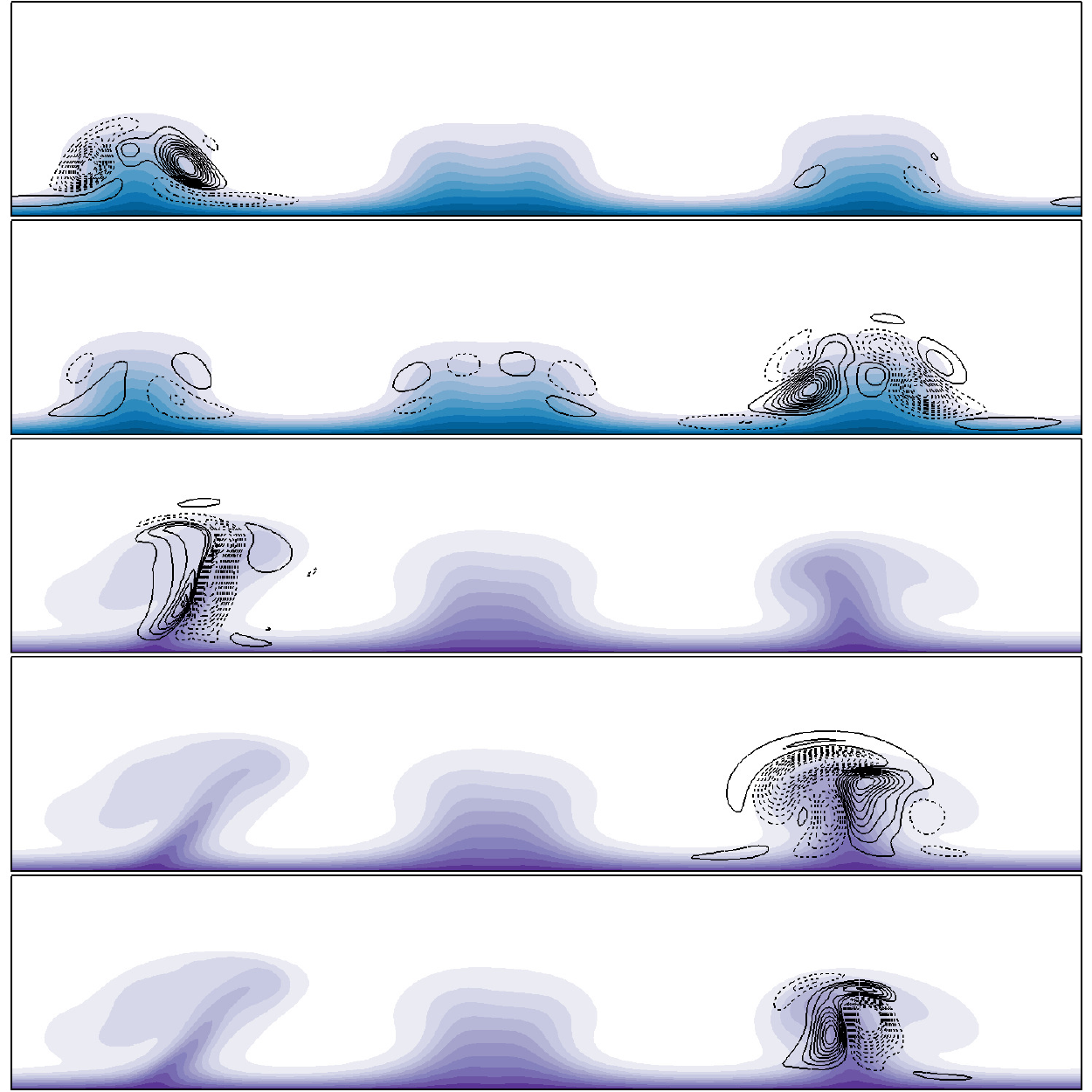}
\end{minipage}
\put(-336,110){({\it a\hspace{1pt}})}
\put(-140,110){({\it b\hspace{1pt}})}
\put(-140,27){({\it c\hspace{1pt}})}
\put(-99,14){$x=50$, Mode 1}
\put(-99,-11){$x=50$, Mode 2}
\put(-99,-36){$x=70$, Mode 1}
\put(-99,-61){$x=70$, Mode 2}
\put(-99,-86){$x=70$, Mode 3}
\put(-290,42){\textcolor{white}{\scriptsize streaks}}
\put(-260,48){\textcolor{white}{\tiny A}}
\put(-260,43){\textcolor{white}{\tiny B}}
\put(-260,38){\textcolor{white}{\tiny C}}
\put(-324,82){\textcolor{black}{$\big\{$}}
\put(-330,82){\rotatebox{0}{$z$}}
\put(-265,42){\textcolor{white}{$\big\{$}}
\put(-290,10){\textcolor{white}{\scriptsize mode 1}}
\put(-273,3){\textcolor{white}{\scriptsize 2, 3}}
\put(-288,94){$x=44$}
\put(-257,94){$50$}
\put(-234,94){$70$}
\put(-198,94){$99$}
\put(-324,33){\rotatebox{0}{$\Bigg\{$}}
\put(-327,-13){\rotatebox{0}{\Large$\Bigg\{$}}
\put(-324,-57){\rotatebox{0}{$\Bigg\{$}}
\put(-334,21){\rotatebox{90}{streaks}}
\put(-344,-30){\rotatebox{90}{secondary}}
\put(-334,-35){\rotatebox{90}{instabilities}}
\put(-344,-95){\rotatebox{90}{helical vortex}}  
\put(-334,-81){\rotatebox{90}{filaments}}
\put(-290,-84){\rotatebox{0}{\textcolor{white}{\scriptsize skeleton mode}}}
\put(-290,59){\rotatebox{0}{\textcolor{black}{\scriptsize shadow mode}}}
\put(-290,81){\scriptsize source}
\put(-327,-98){$x$:}
\put(-120,-100){\scriptsize A}
\put(-88,-100){\scriptsize B}
\put(-46,-100){\scriptsize C}
\end{center}
\caption{(a) Image-based flow decomposition of boundary layer transition. $M_x=M_z=3$, leading to nine EWT modes. Important streamwise locations, $x=44$ (onset of secondary instabilities), $x=50,~70$ (base flows and eigenfunctions as seen in panel (c)), $x=99$ (formation of hairpin vortices) are highlighted. White dashed lines in $\mathcal{W}_{2,1}$, $\mathcal{W}_{2,2}$, $\mathcal{W}_{2,3}$ show the position of streaks and the streamwise range over which secondary instabilities occur. (b) Growth rates of the secondary instabilities versus streamwise wavenumber $\alpha$ at cross sections of $x=43,~44,~45,~50,~70$. The red cross marker indicates the maximum growth rate of a mode whose eigenfunction is visualised in panel (c). The eigenfunction $u^\prime$ is shown with black lines (dashed lines denote negative values) on top of the base flow contours ($u=0.1, 0.2, ...,0.9$). A movie of the video-based flow decomposition is available as a supplementary file.}
\label{F3_bypass2}
\end{figure}

We show the flow decomposition in Figure~\ref{F3_bypass2}(a). In this example, we have $M_x=M_z=3$. By construction, mode ${\mathcal{W}_{1,1}}$ (shadow mode) and ${\mathcal{W}_{3,3}}$ (skeleton mode) represent the flow structure with a spectrum that amounts to the lower- and higher- end both in the streamwise and spanwise directions. Mode ${\mathcal{W}_{1,2}}$ and ${\mathcal{W}_{1,3}}$ isolate the streaks. The secondary instabilities of streaks are captured by modes ${\mathcal{W}_{2,1}}$, ${\mathcal{W}_{2,2}}$ and ${\mathcal{W}_{2,3}}$, featuring the low-to-high wave numbers in the spanwise direction. From ${\mathcal{W}_{2,1}}$, it is seen that Streak A gives rise to sinuous instabilities after $x=40$, followed by the instability of Streak C. It also indicates that perturbations around Streak C have a larger amplitude after $x=70$. These observations are however hidden in the source image. Finally, in mode ${\mathcal{W}_{3,1}}$ and ${\mathcal{W}_{3,2}}$, localised helical vortex filaments are identified, marking the meandering of streaks and appearance of smaller scales.

To verify the observation from EWT modes, a biGlobal stability analysis (secondary instability of streaks) is performed at $x=43,44,45$ (to secure the onset location of secondary instabilities) and $x=50,70$ (to identify multiple modes). We plot temporal growth rates versus streamwise wavenumber $\alpha$ in panel (b). Their peak growth rates are marked with red crosses, and the corresponding eigenfunctions \& base flows are provided in panel (c). A positive growth rate is seen at $x=44$, which demonstrates the onset of secondary instabilities. This mode (termed mode 1) has its eigenfunctions localised around Streak A. Further downstream, the flow is unstable to multiple modes. From around $x=50$, mode 2 (centred around Streak C) also becomes unstable, though, at a smaller growth rate. However, further downstream at $x=70$, mode 3 (centred around Streak C) becomes the most energetic. All these secondary instability modes are of sinuous nature. Generally, the stability analysis matches rather well with EWT modes; the onset of multiple secondary instability modes, their sinuous nature and the dominance of instabilities around Streak C are evidenced by mode ${\mathcal{W}_{2,1}}$. For the case of a video input, an unsteady flow decomposition is provided as a supplementary file. The video records the flow from $t=0$ when FST is introduced to the laminar flow until the time step illustrated in Figure~\ref{F2_bypass1} and \ref{F3_bypass2}. The video-based EWT modes provide the evolution of flow structures subject to the temporal-spatial averaged Fourier supports.

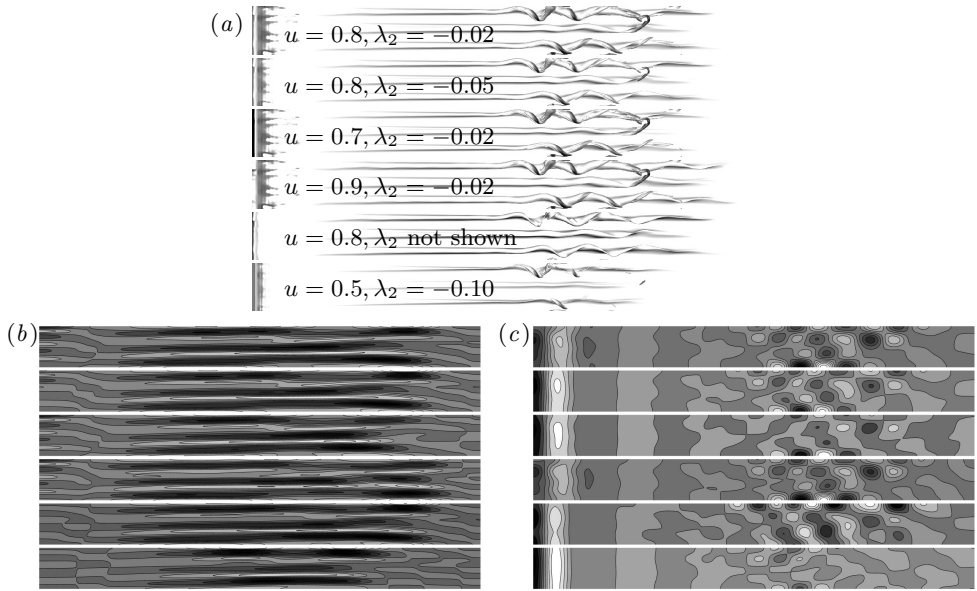
\begin{figure}
\begin{center}
\includegraphics[width=0.5\linewidth]{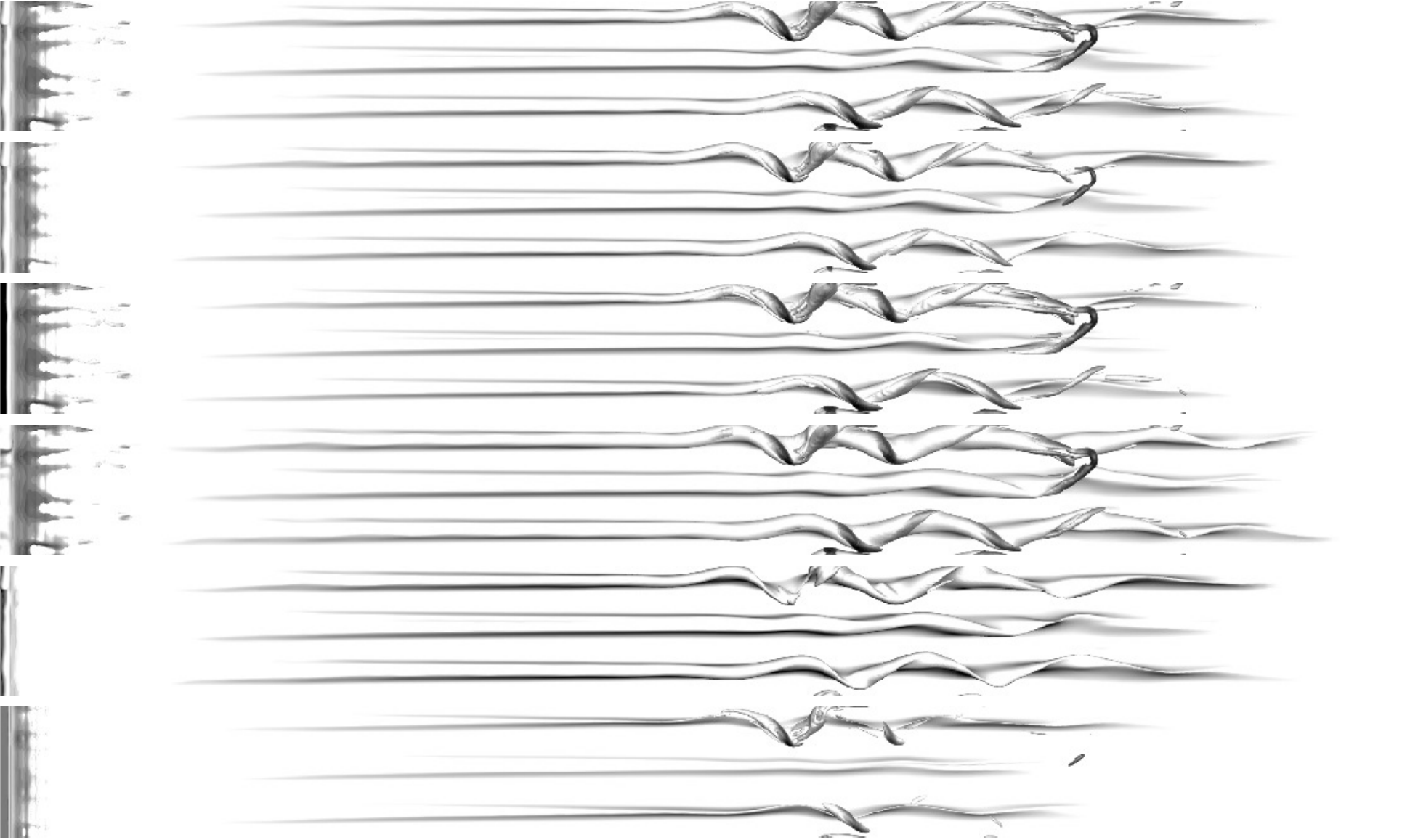}
\put(-208,105){({\it a\hspace{1pt}})}
\put(-180,101){$u=0.8, \lambda_{2}=-0.02$}
\put(-180,82){$u=0.8, \lambda_{2}=-0.05$}
\put(-180,63){$u=0.7, \lambda_{2}=-0.02$}
\put(-180,44){$u=0.9, \lambda_{2}=-0.02$}
\put(-180,25){$u=0.8, \lambda_{2}$ not shown}
\put(-180,6){$u=0.5, \lambda_{2}=-0.10$}
\\
\includegraphics[width=0.45\linewidth]{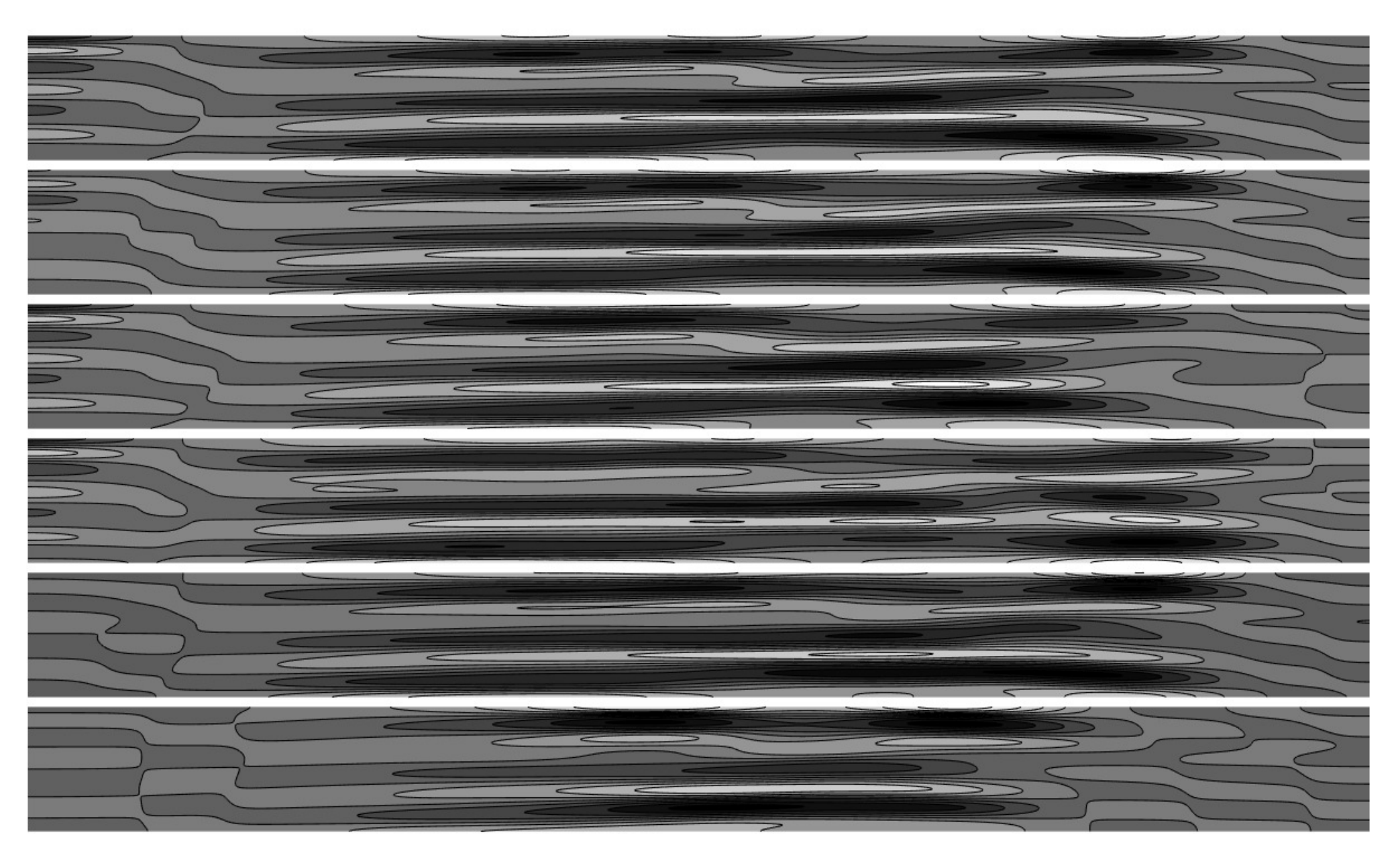}\hspace{10pt}
\includegraphics[width=0.45\linewidth]{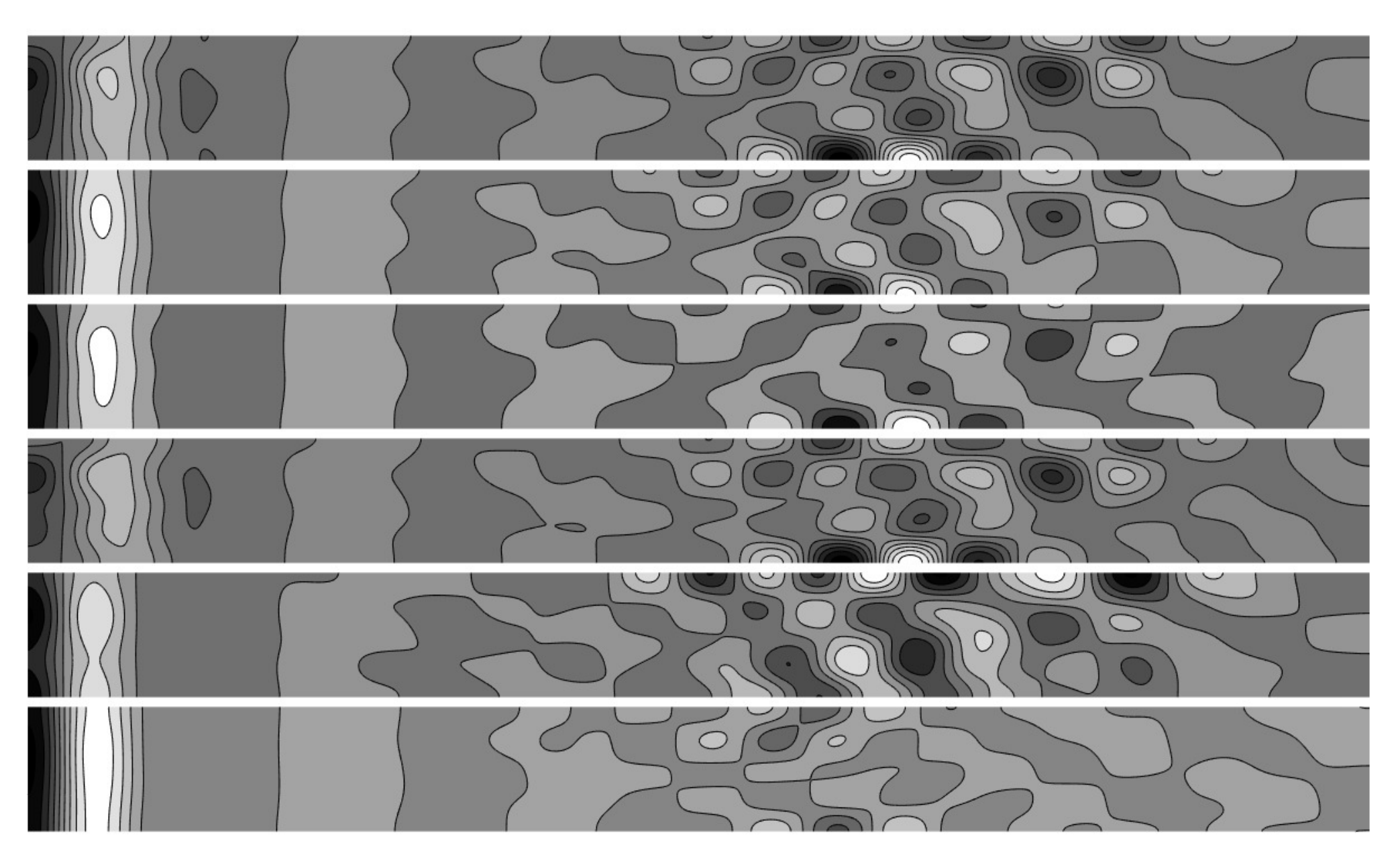}
\put(-368,94){({\it b\hspace{1pt}})}
\put(-183,94){({\it c\hspace{1pt}})}
\end{center}
\caption{\RR (a) Flow visualisations of boundary layer transition with iso-surfaces of $u$ and $\lambda_{2}$ at values indicated on the image. (b) and (c) The corresponding EWT modes: $\mathcal{W}_{1,2}$ and $\mathcal{W}_{2,1}$.}
\label{F_bypass_robus}
\end{figure}

{\RR To test the sensitivity and robustness of the method with respect to the input image, a comparison of EWT modes with different flow visualisations is shown in figure~\ref{F_bypass_robus}.  Images with various isosurfaces of $u$ and $\lambda_{2}$ are adopted as shown in panel (a). 
The same filter banks from EWT are used to process these inputs. As can be seen from panel (b) and (c), the streaks and their secondary instabilities are correctly extracted by mode $\mathcal{W}_{1,2}$ and $\mathcal{W}_{2,1}$ for most of the cases, regardless of the different inputs. With isosurfaces of $u$ alone, secondary instabilities are captured, but the importance of Streak C is not revealed. For a more inimical input ($u=0.5, \lambda_{2}=-0.10$), the EWT modes still result in satisfactory decompositions.}

\begin{figure}
\begin{center}
\includegraphics[width=0.32\linewidth]{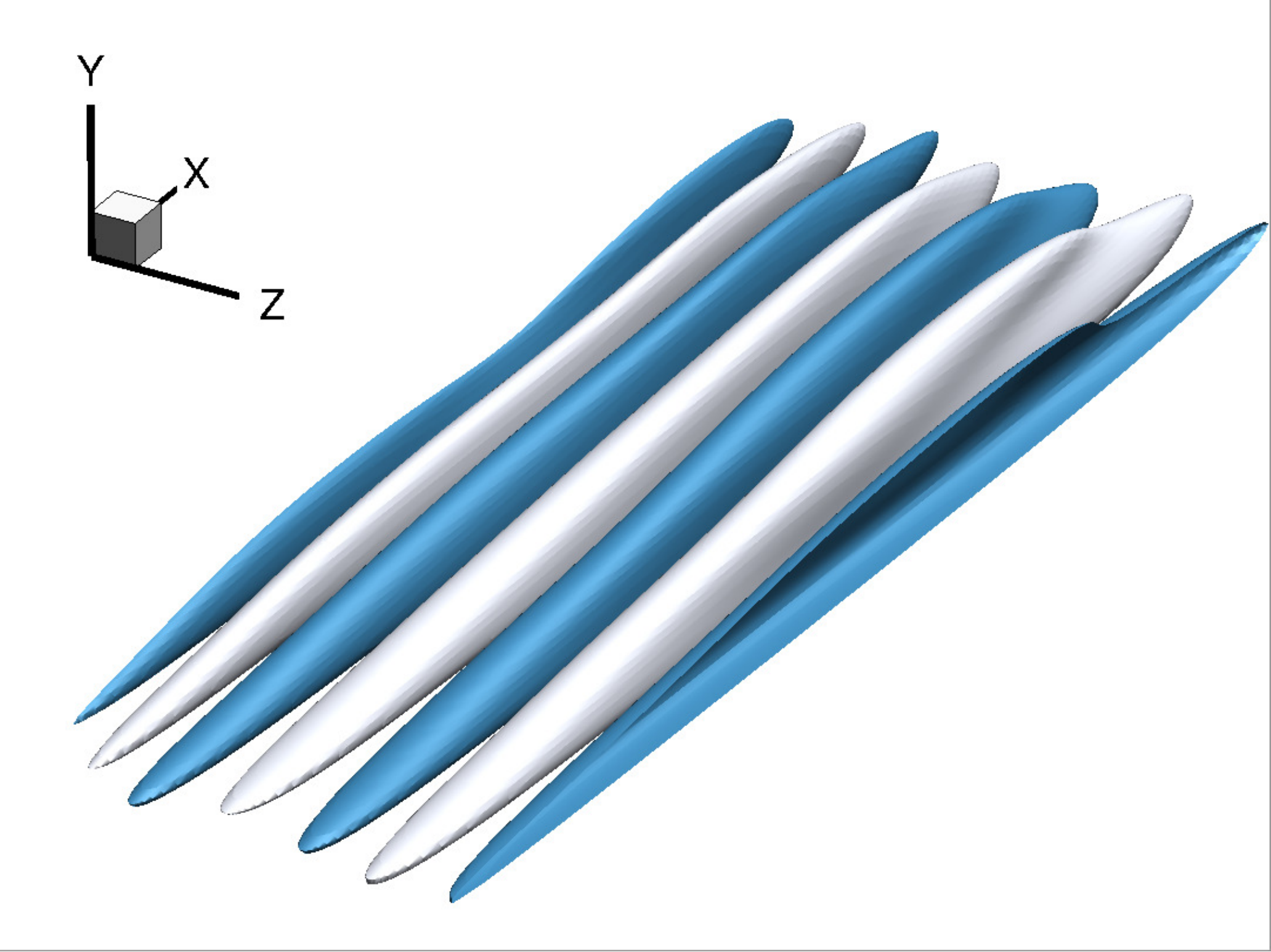}
\includegraphics[width=0.32\linewidth]{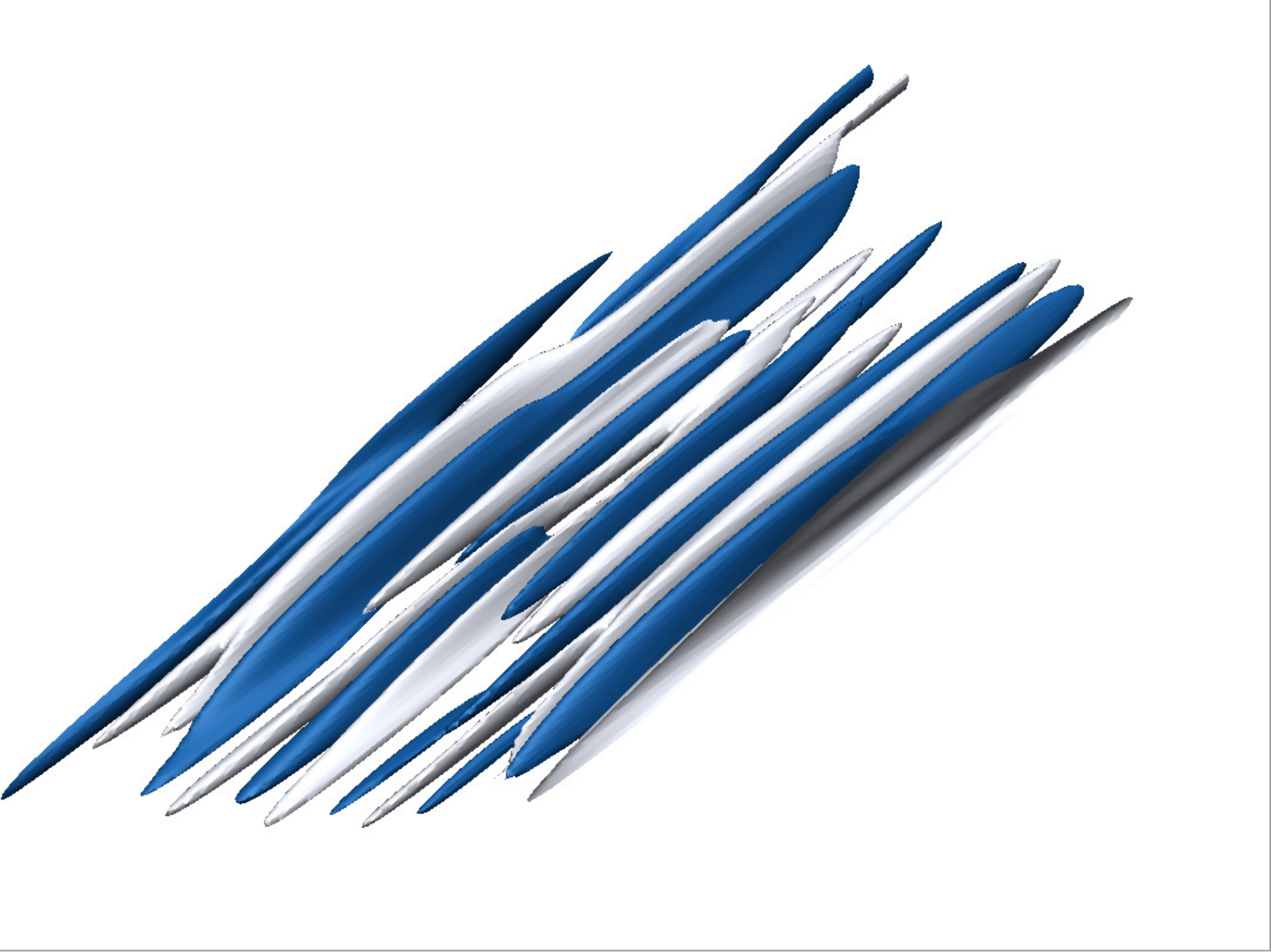} 
\includegraphics[width=0.32\linewidth]{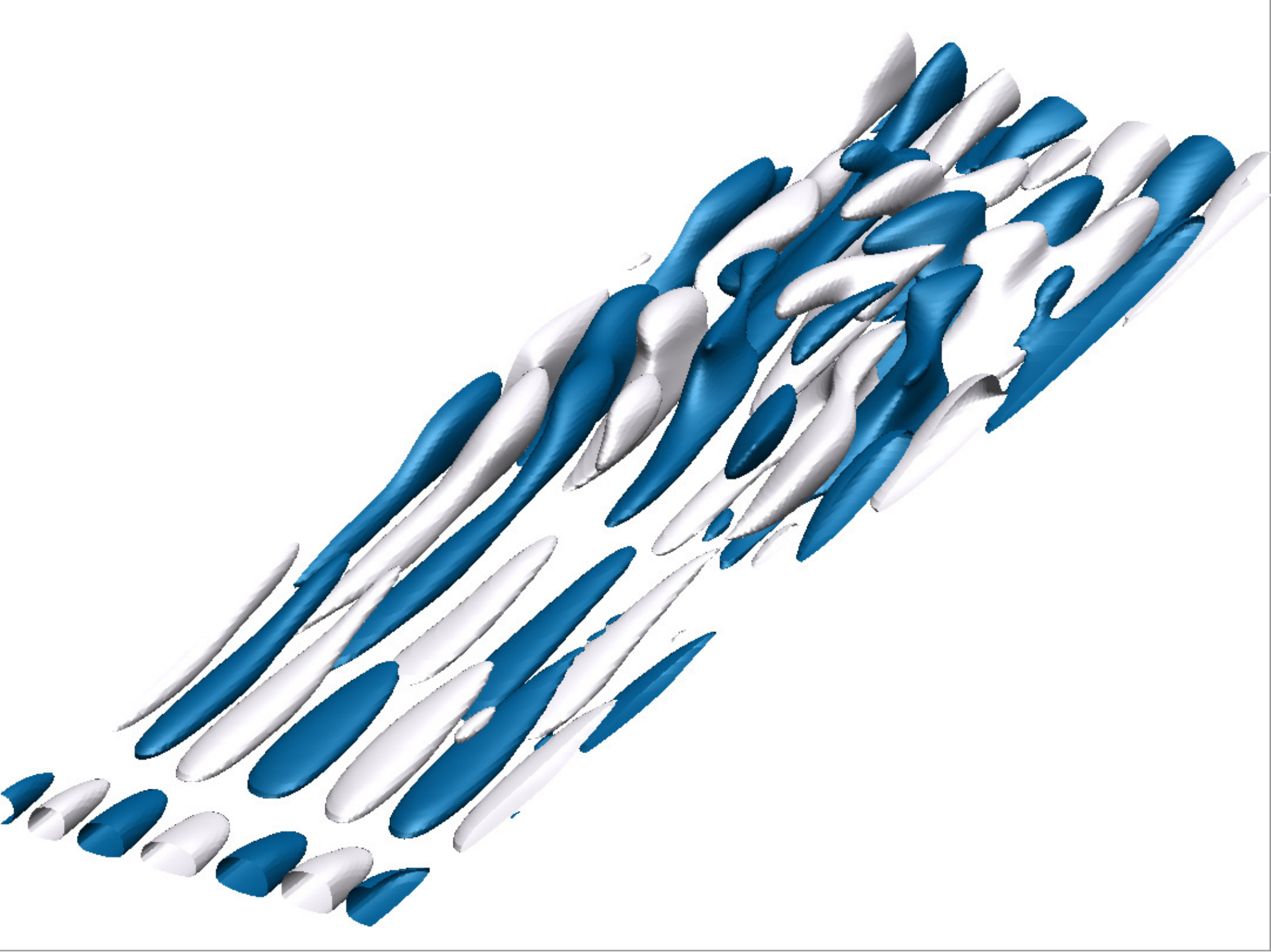}
\put(-290,25){$\mathcal{W}_{1,2}$}
\put(-170,25){$\mathcal{W}_{1,3}$}
\put(-40,25){$\mathcal{W}_{2,2}$}
\\
\includegraphics[width=0.32\linewidth]{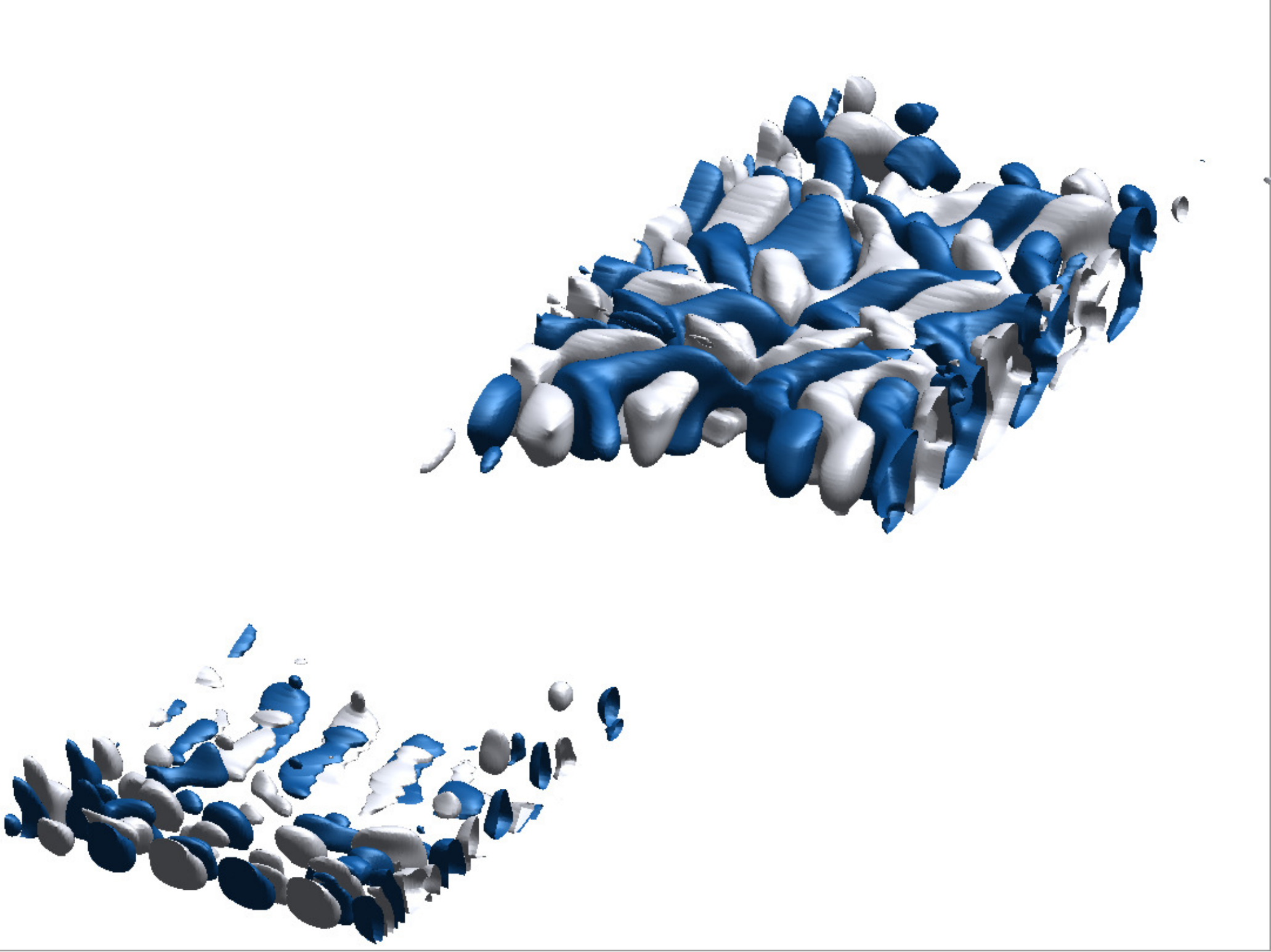}
\includegraphics[width=0.32\linewidth]{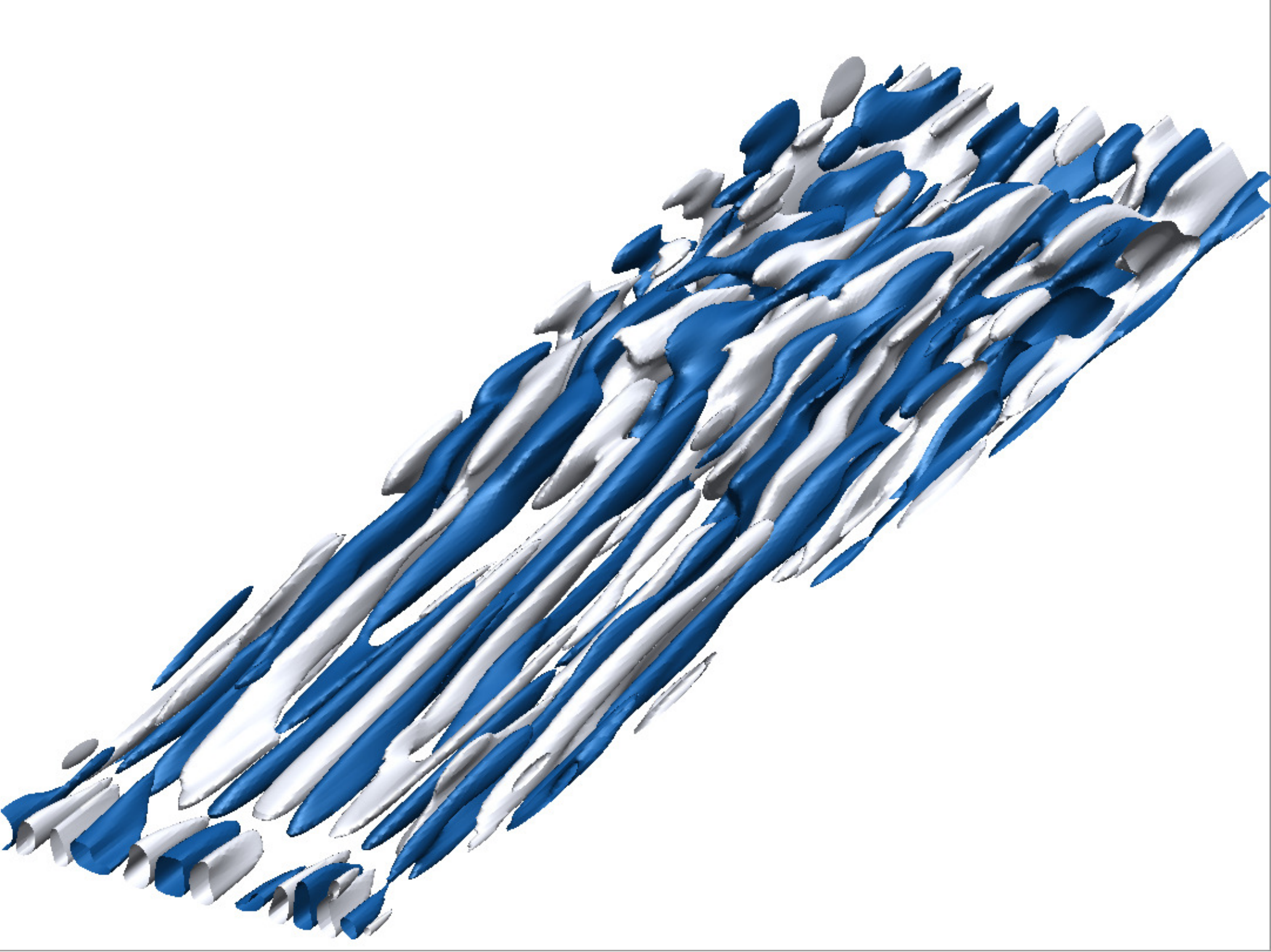}
\includegraphics[width=0.32\linewidth]{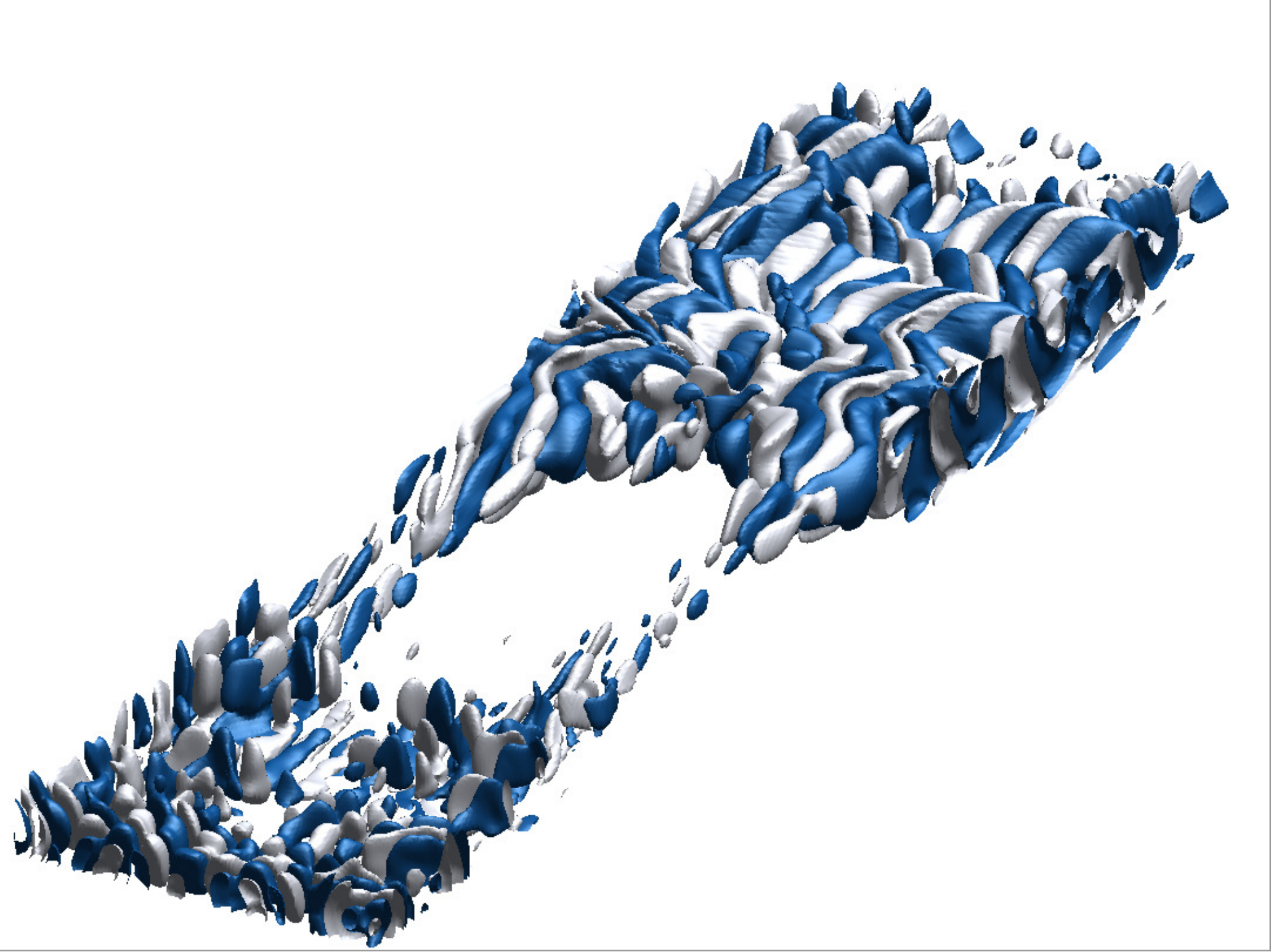}
\put(-290,25){$\mathcal{W}_{3,2}$}
\put(-170,25){$\mathcal{W}_{2,3}$}
\put(-40,25){$\mathcal{W}_{3,3}$}
\end{center}
\caption{\RR Flow decomposition with 3D velocity data $u(x,y,z)$. EWT is applied in the $x$ and $z$ directions with  $M_x=M_z=3$. Iso-surfaces are defined and coloured (blue/white for positive/negative values) according to EWT modes: $\mathcal{W}_{1,2}$ ($\pm0.1$), $\mathcal{W}_{1,3}$ ($\pm0.05$), $\mathcal{W}_{2,2}$ ($\pm0.02$), $\mathcal{W}_{2,3}$ ($\pm0.01$), $\mathcal{W}_{3,2}$ ($\pm0.02$),  $\mathcal{W}_{3,3}$ ($\pm0.01$).}
\label{F_bypass5}
\end{figure}

{\RR We further apply EWT to the decomposition of 3D instantaneous data, as shown in figure~\ref{F_bypass5}. The flow has been decomposed in the $x$ and $z$ directions with $M_x=M_z=3$ based on the streamwise velocity $u(x,y,z)$ on each $y$ plane. We show the physically significant modes with iso-surfaces. As can be seen, the streamwise-elongated streaks and their nonlinear harmonics are extracted by $\mathcal{W}_{1,2}$ and $\mathcal{W}_{1,3}$. $\mathcal{W}_{2,2}$ and $\mathcal{W}_{3,2}$ capture streak instabilities at different scales.  $\mathcal{W}_{2,3}$ and $\mathcal{W}_{3,3}$ stand for the higher-end spectrum in the spanwise direction, displaying the canonical bypass nature of the transition process, \ie,  small scales in the freestream only reappear further downstream in the boundary layer leading to a `clean area'. Comparing the results presented in figure~\ref{F3_bypass2} and \ref{F_bypass5}, both inputs of the instantaneous 3D data and its visualisation leads to adaptively and physically-relevant decompositions. The extracted streaks and their secondary instabilities from both results match well with each other and correctly reflect the flow physics. Note that when viewed and processed as an image, flow visualisation stands for a projection of the 3D instantaneous data, with the objective to compress while retaining critical features of the data. Differences, as a result, stem from the discrepancy in the Fourier supports.  For example, secondary instabilities are extracted by $\mathcal{W}_{2,1}$, $\mathcal{W}_{2,2}$ and $\mathcal{W}_{2,3}$ from the visualisation, while  $\mathcal{W}_{2,2}$ \& $\mathcal{W}_{3,2}$ from 3D instantaneous data stand for streak instabilities. Another advantage of using flow data is that the extracted modes will have a solid physical meaning. For example, in this case, the modes are the decomposed streamwise velocities. 3D data further leads to more details of the flow, as evidenced by the bypass characteristics in $\mathcal{W}_{2,3}$ and $\mathcal{W}_{3,3}$.}

\section{\RR Comparison with the other methods}\label{S41}
{\RR At this point, it becomes merited to ask, what is the added value of EWT modes to the state-of-art data-based flow decompositions (\eg the POD and DMD families)? This section takes the boundary layer transitional flow as an example to show the strengths and weaknesses of the proposed methods. }
\begin{figure}
\begin{center}
\includegraphics[width=0.38\linewidth]{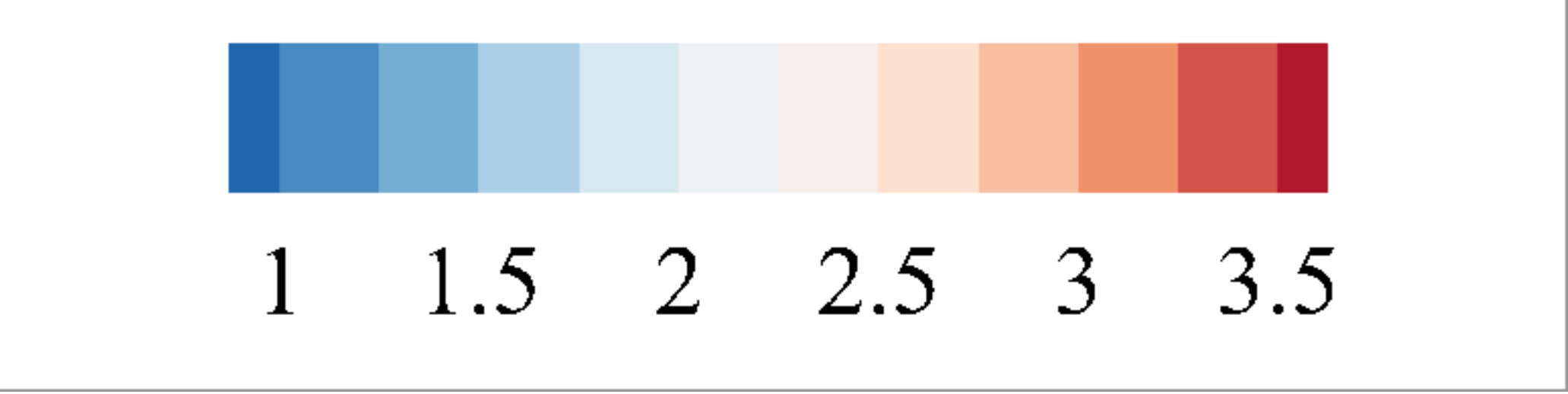}
\put(-136,8.5){$y$: }
\\
\includegraphics[width=0.24\linewidth]{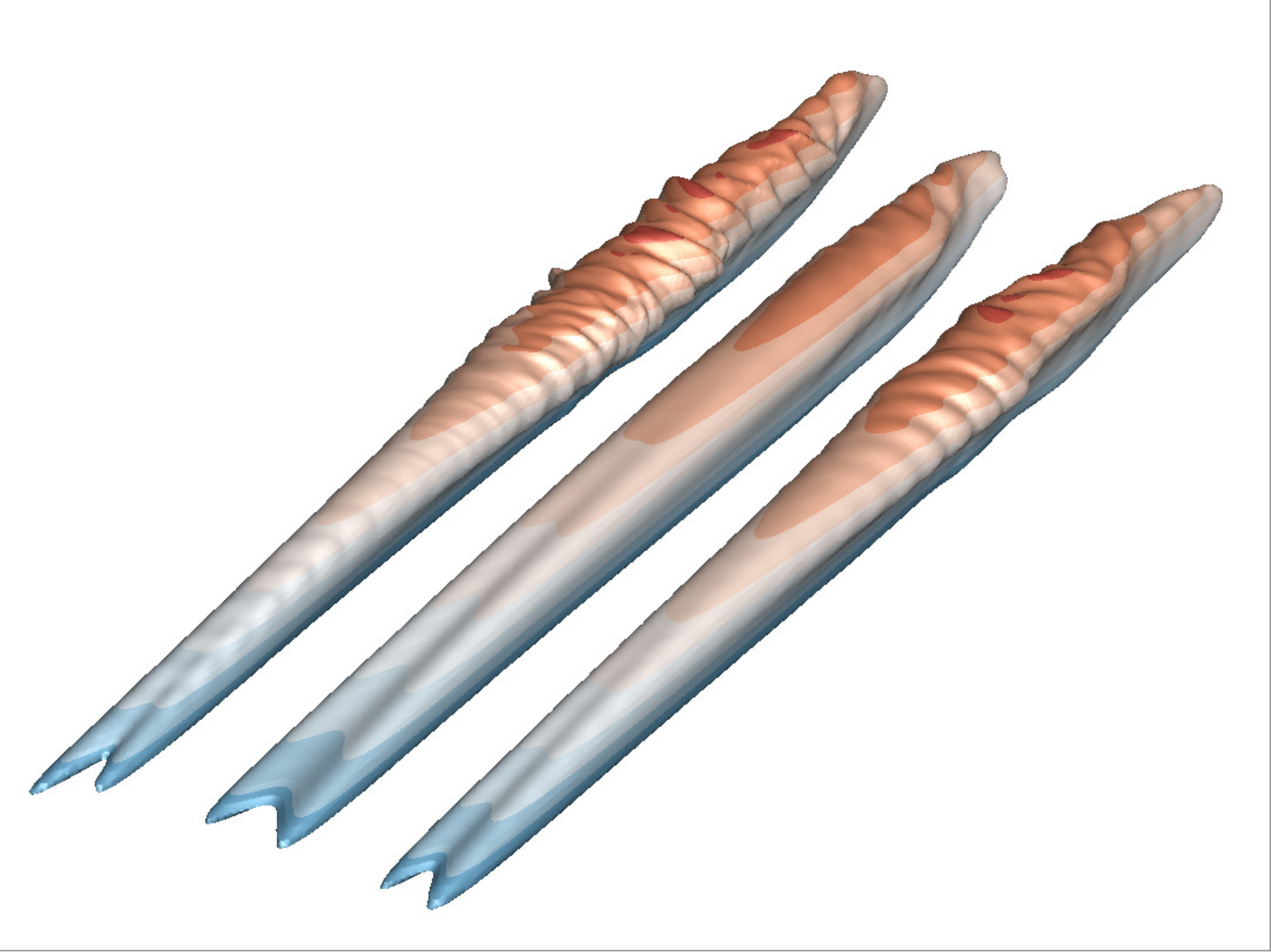}
\includegraphics[width=0.24\linewidth]{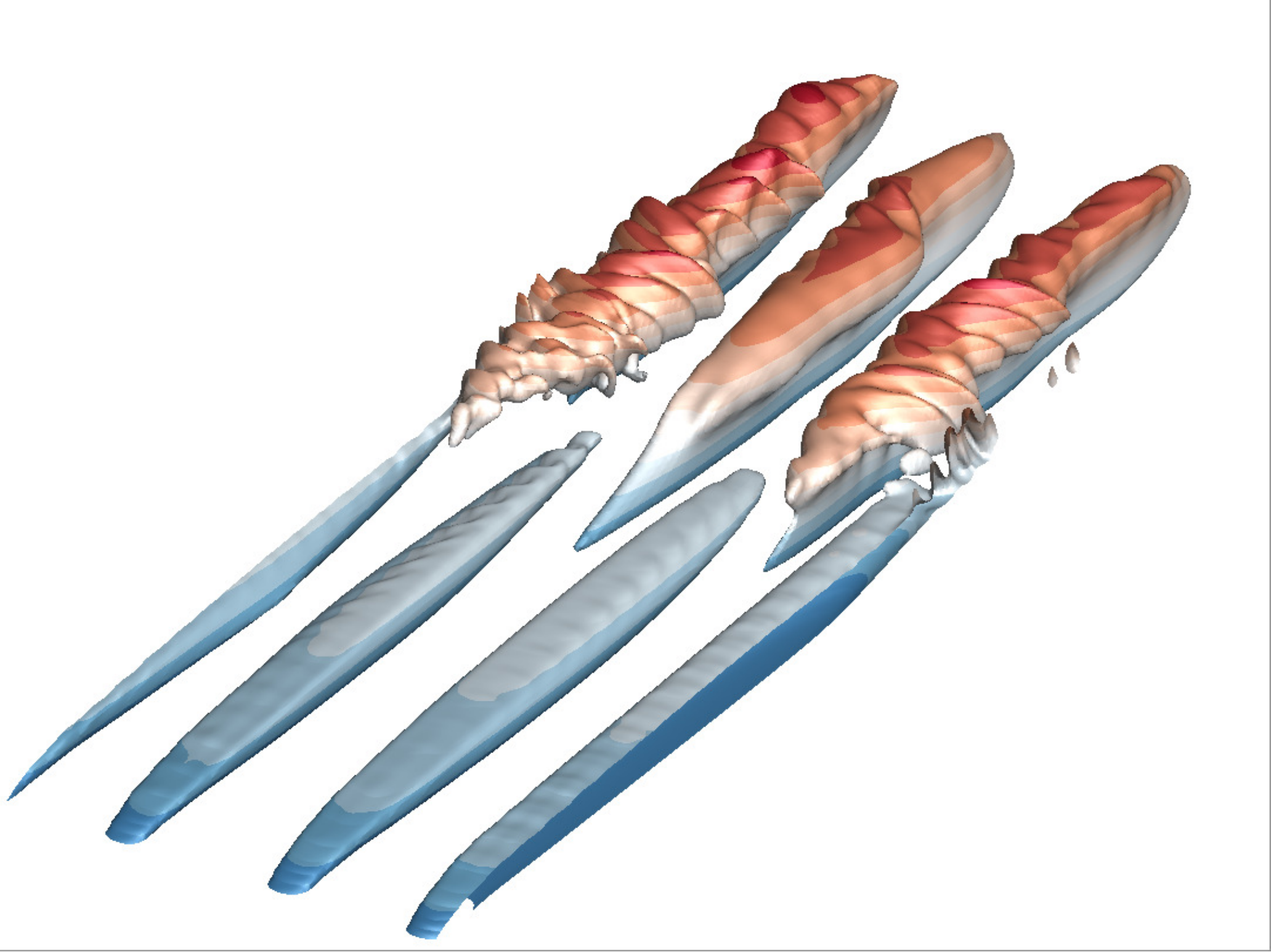} 
\includegraphics[width=0.24\linewidth]{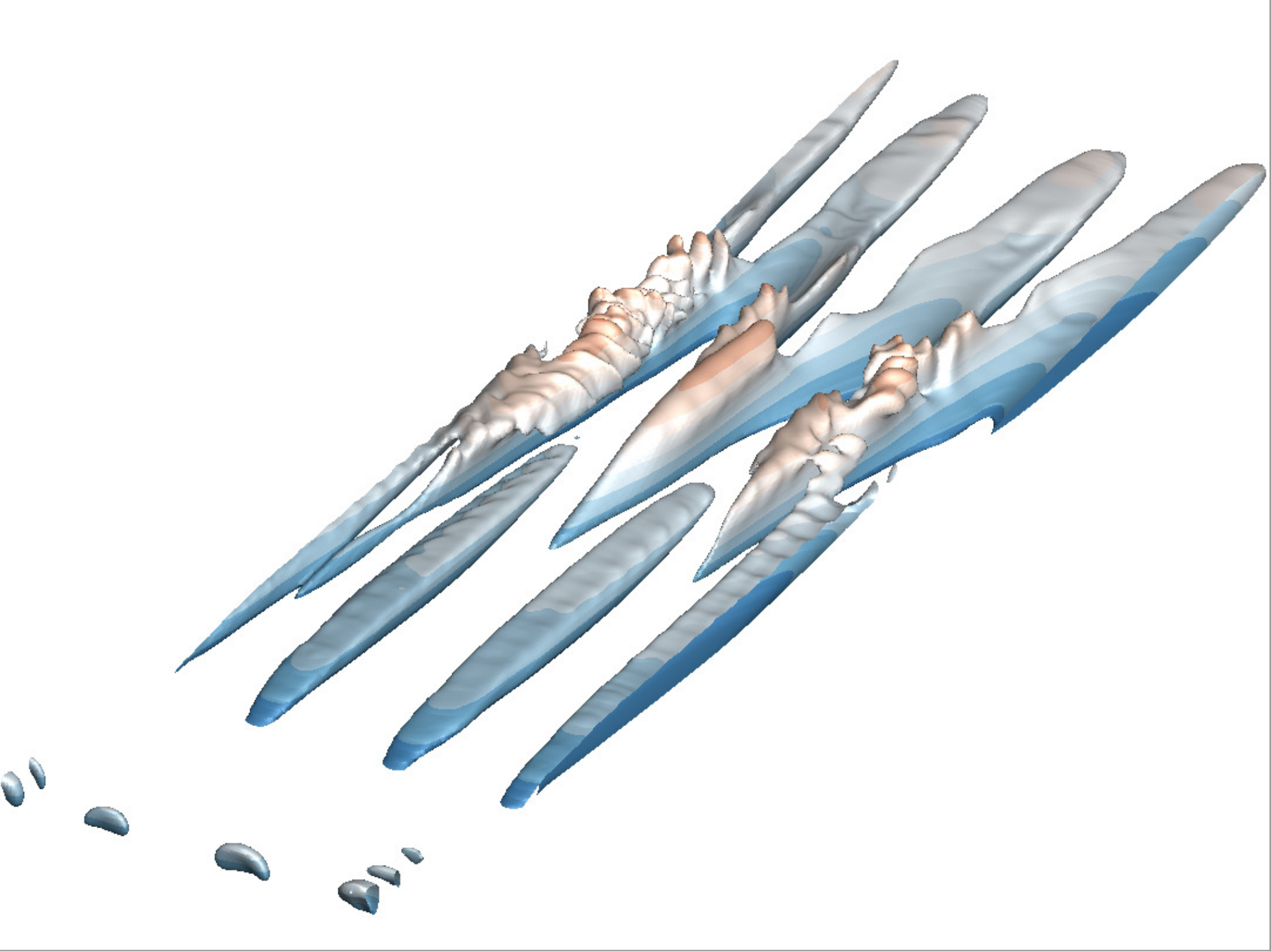}
\includegraphics[width=0.24\linewidth]{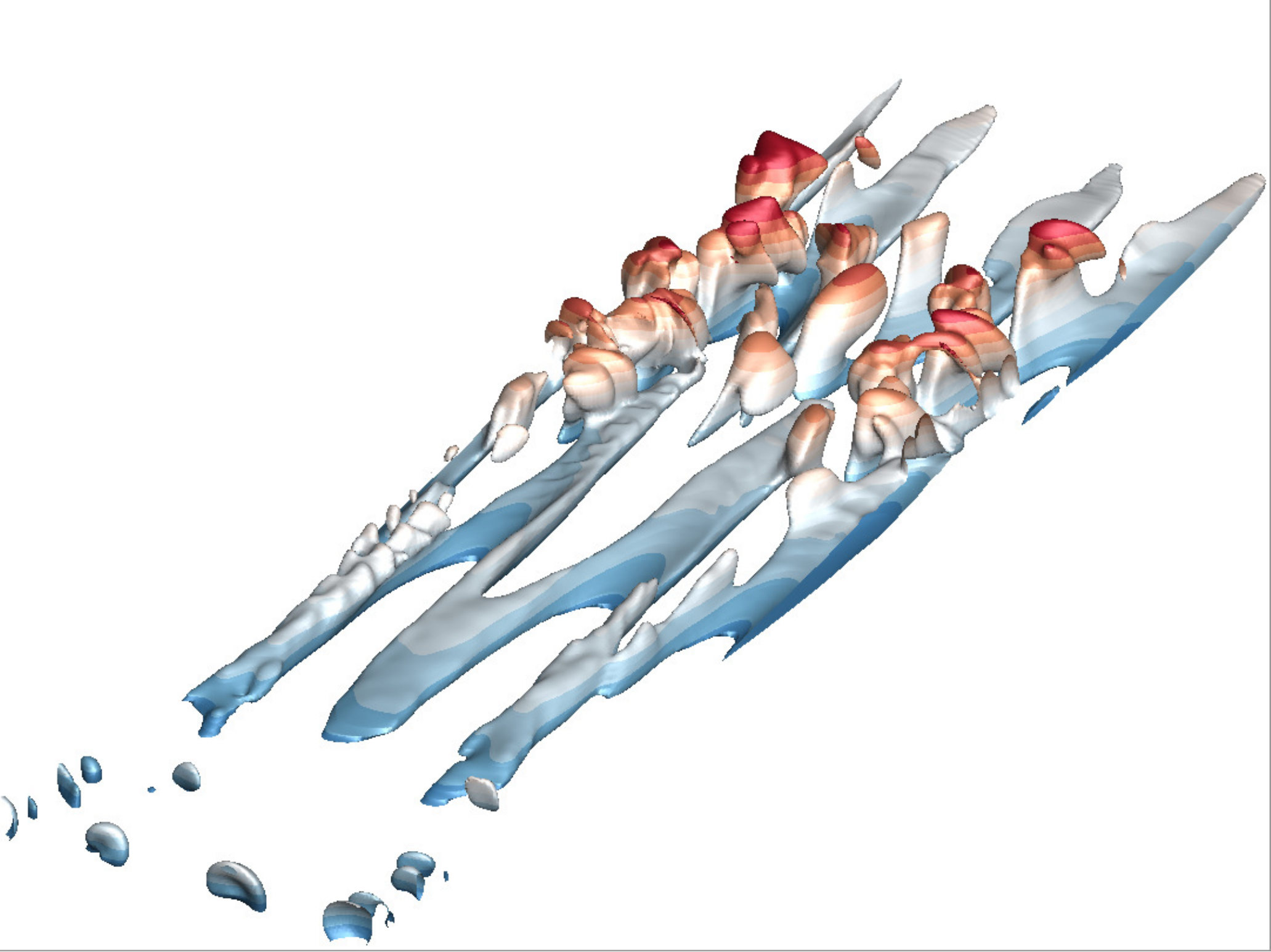}
\put(-375,60){({\it a\hspace{1pt}})}
\put(-312,14){POD}
\put(-322,4){Mode 1}
\put(-201,4){2}
\put(-106,4){3}
\put(-11,4){4}
\\
\includegraphics[width=0.24\linewidth]{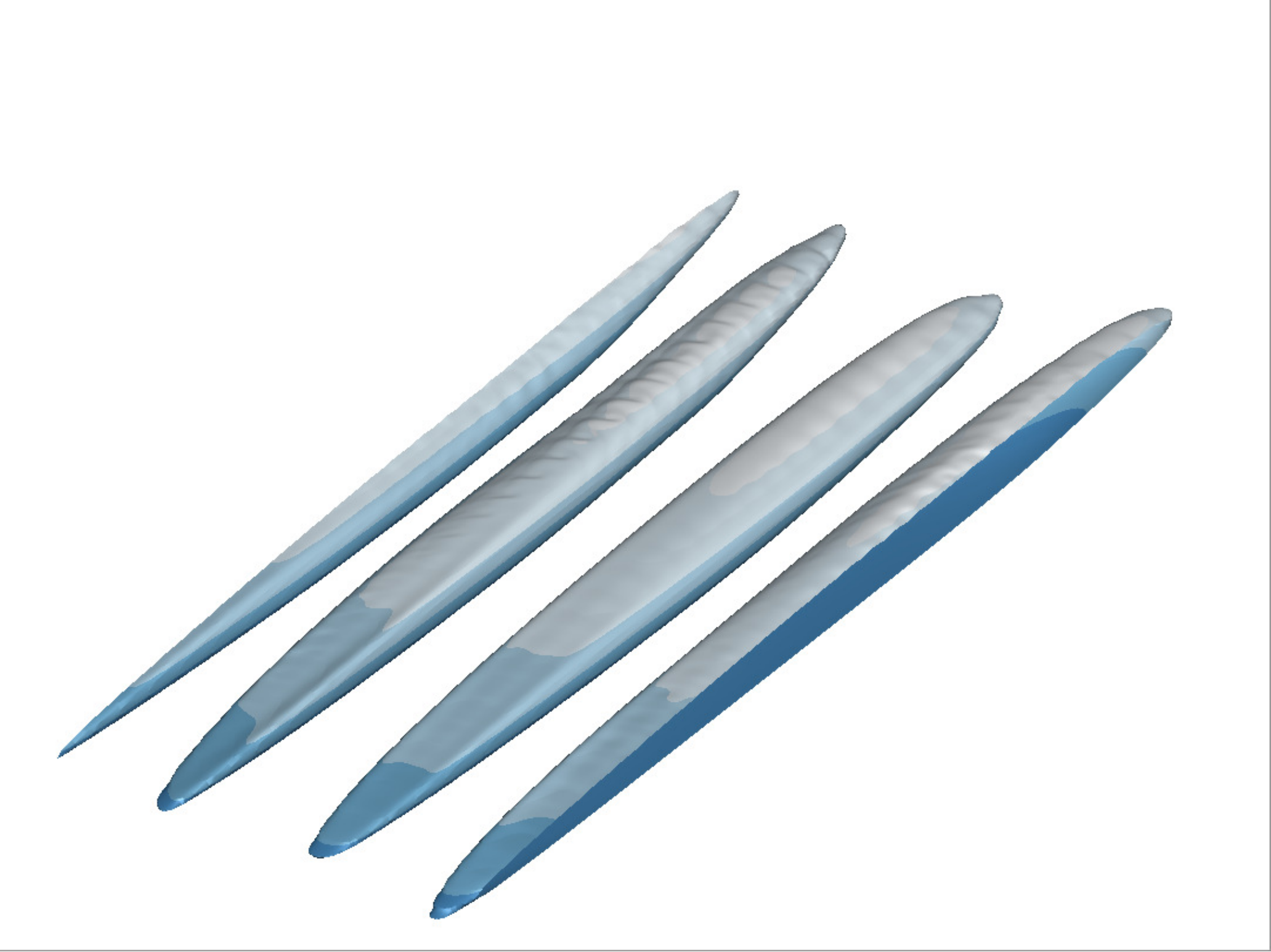}
\includegraphics[width=0.24\linewidth]{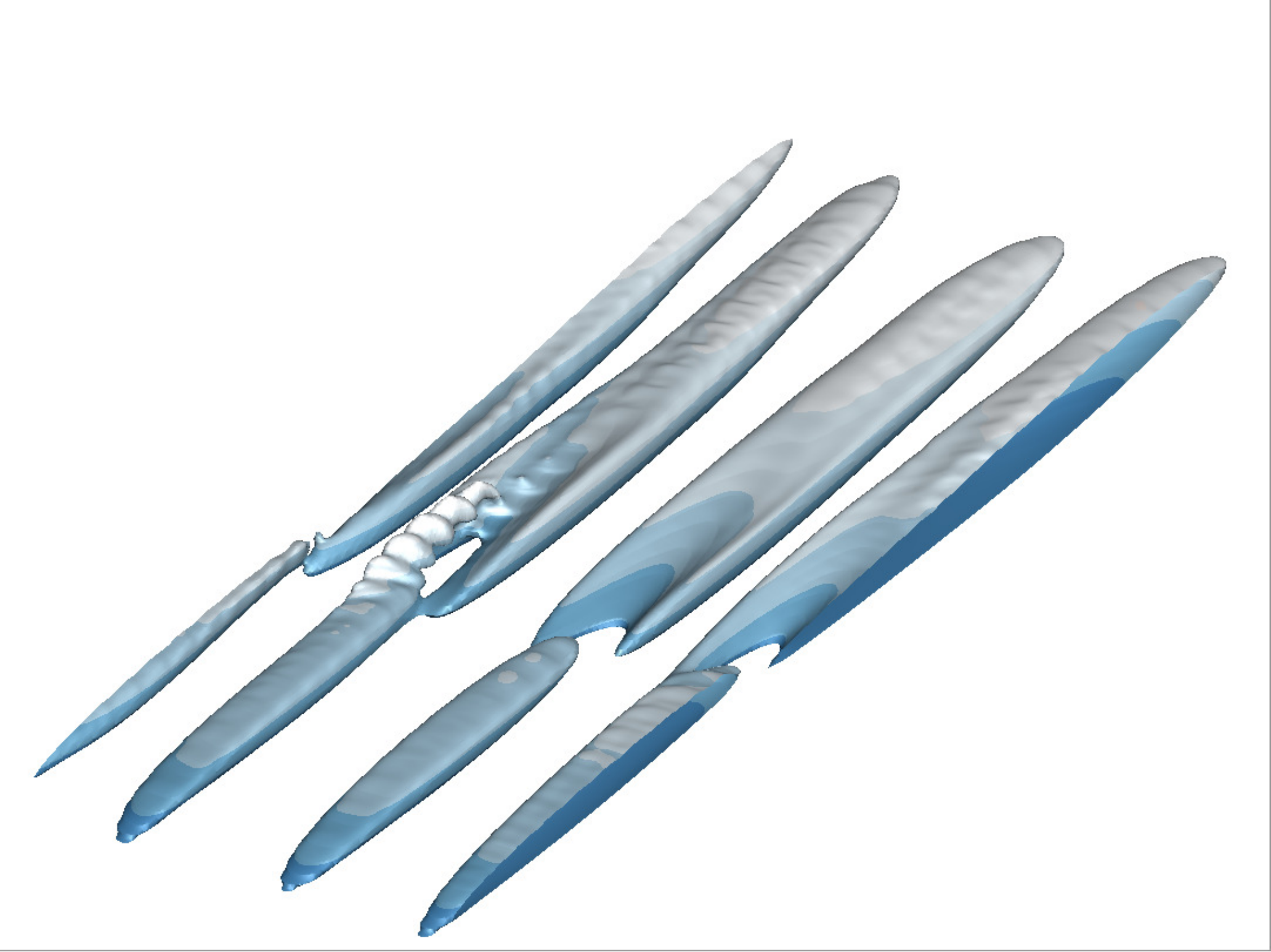} 
\includegraphics[width=0.24\linewidth]{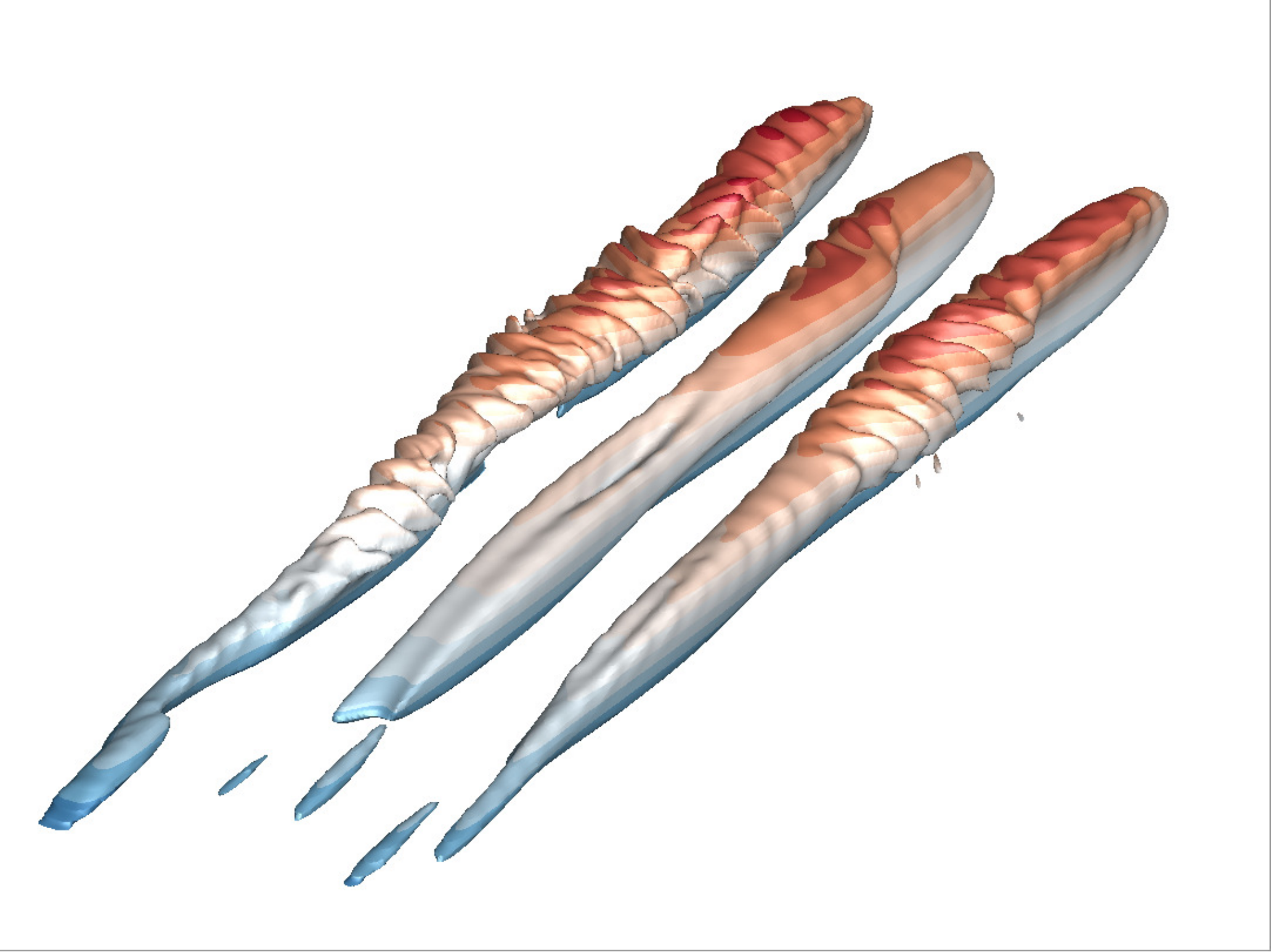}
\includegraphics[width=0.24\linewidth]{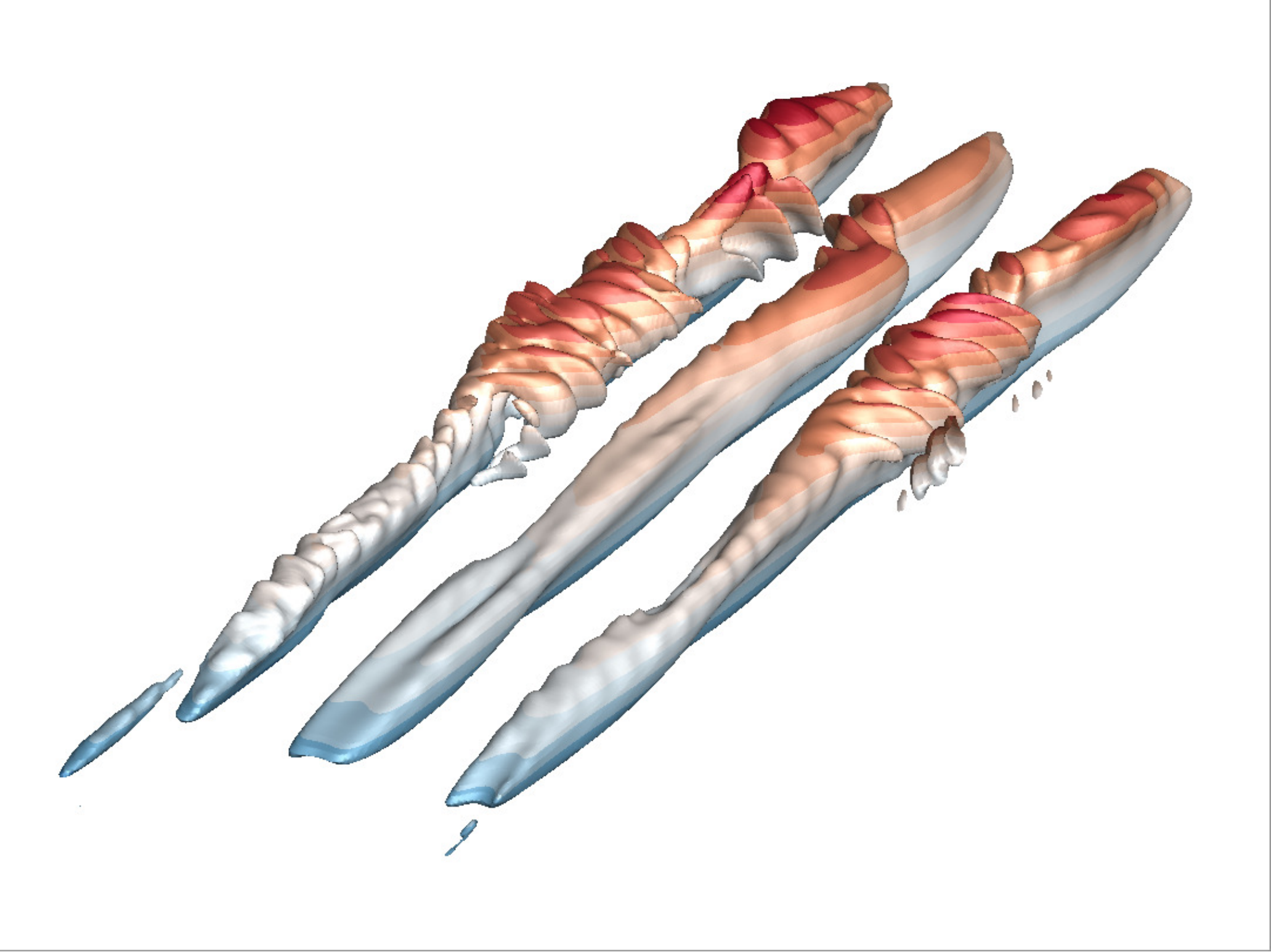}
\put(-375,60){({\it b\hspace{1pt}})}
\put(-313,14){DMD}
\put(-331,4){Mode 1/2}
\put(-209,4){3/4}
\put(-114,4){5/6}
\put(-19,4){7/8}
\\
\includegraphics[width=0.24\linewidth]{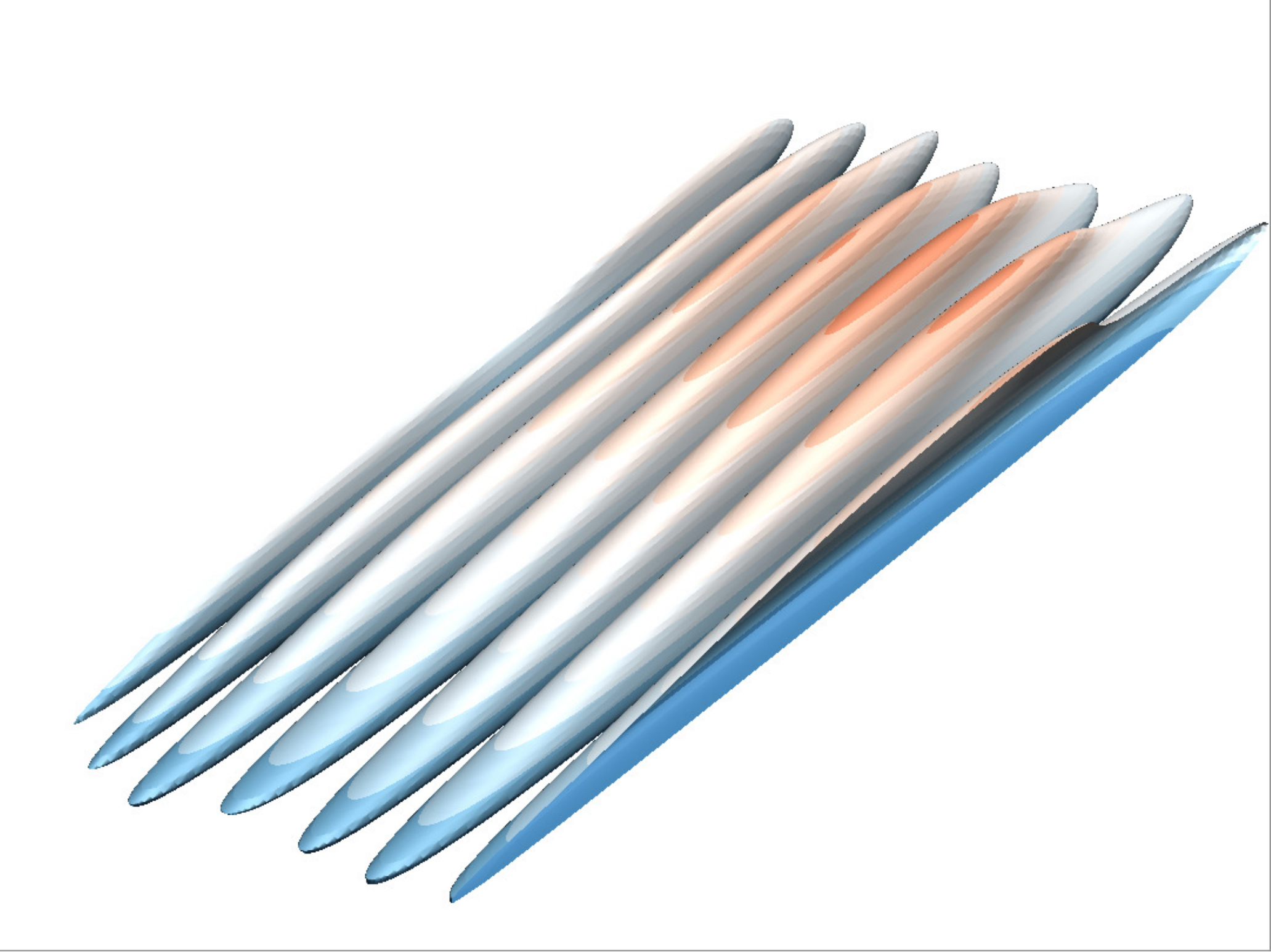}
\includegraphics[width=0.24\linewidth]{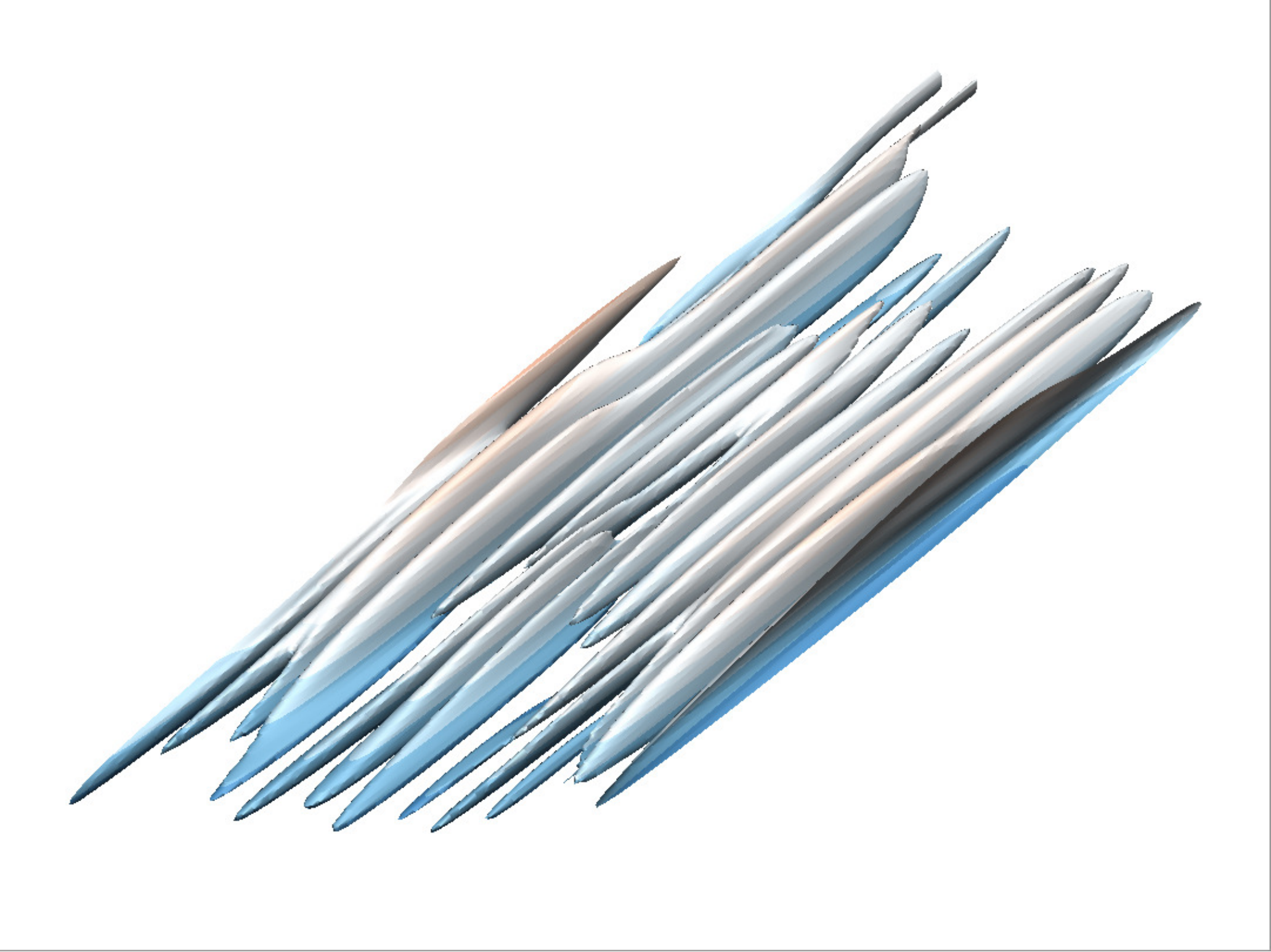} 
\includegraphics[width=0.24\linewidth]{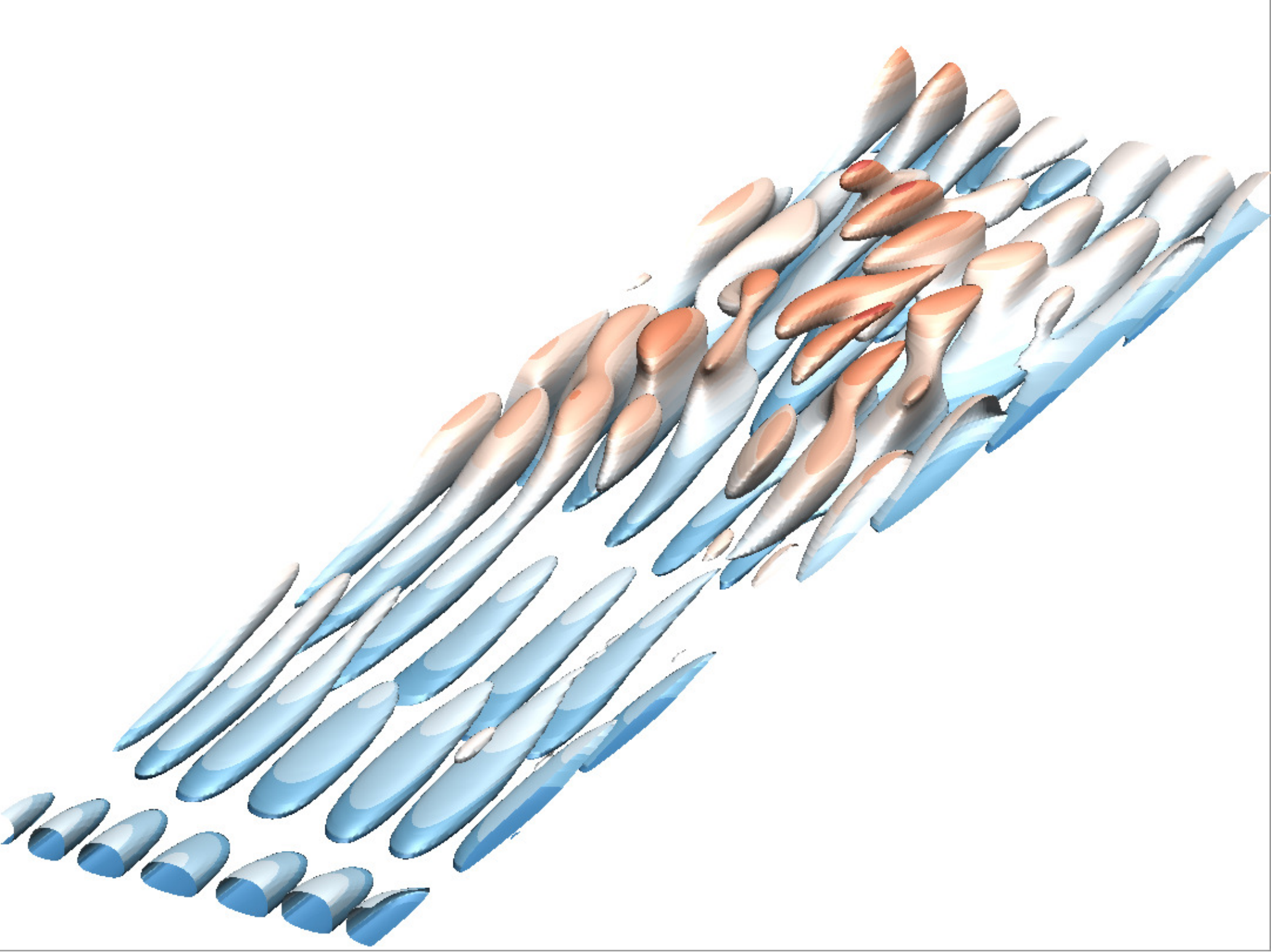}
\includegraphics[width=0.24\linewidth]{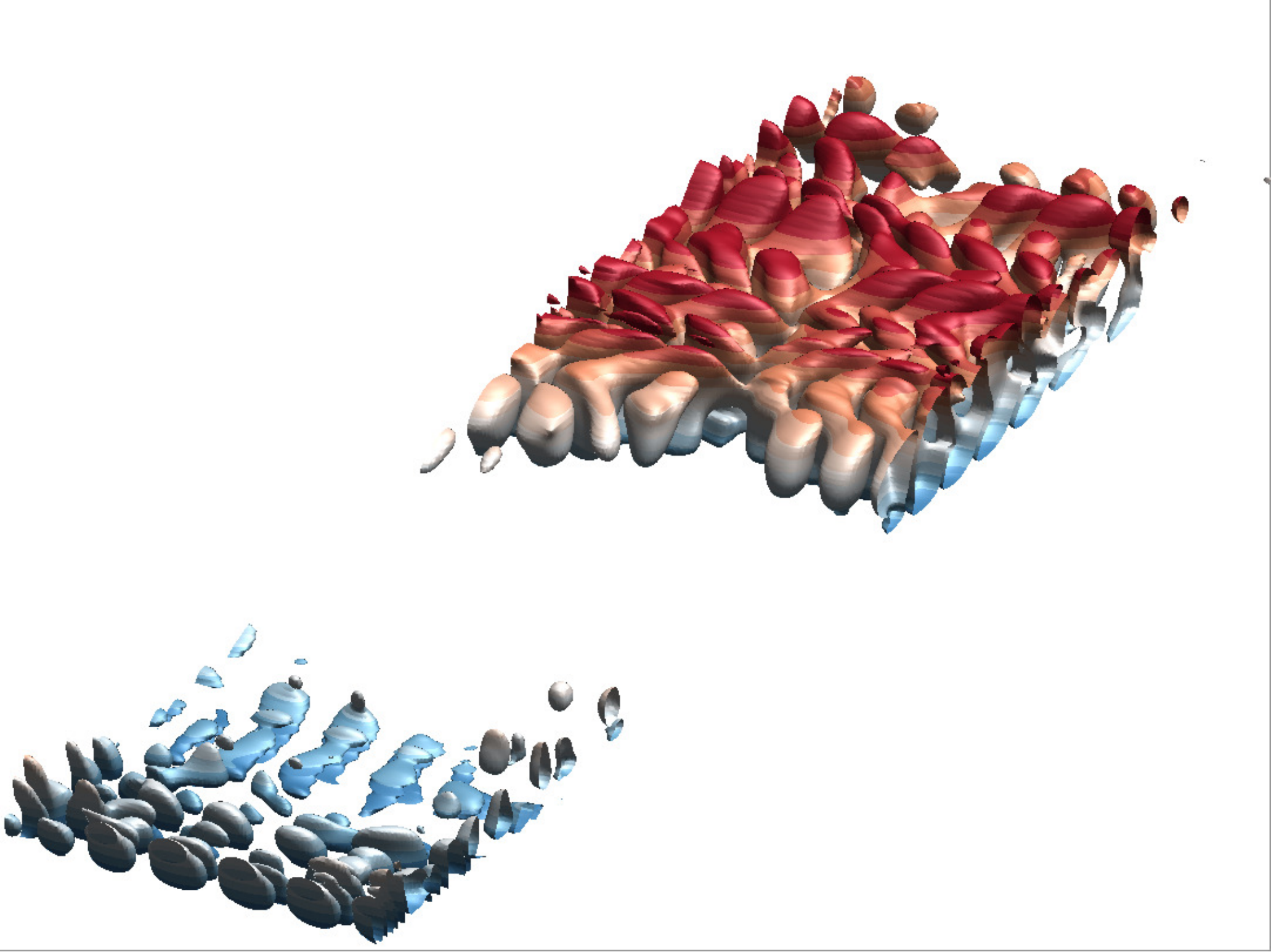}
\put(-375,60){({\it c\hspace{1pt}})}
\put(-331,4){EWT $\mathcal{W}_{1,2}$}
\put(-217,4){$\mathcal{W}_{1,3}$}
\put(-121,4){$\mathcal{W}_{2,2}$}
\put(-26,4){$\mathcal{W}_{2,3}$}
\\
\includegraphics[width=0.32\linewidth]{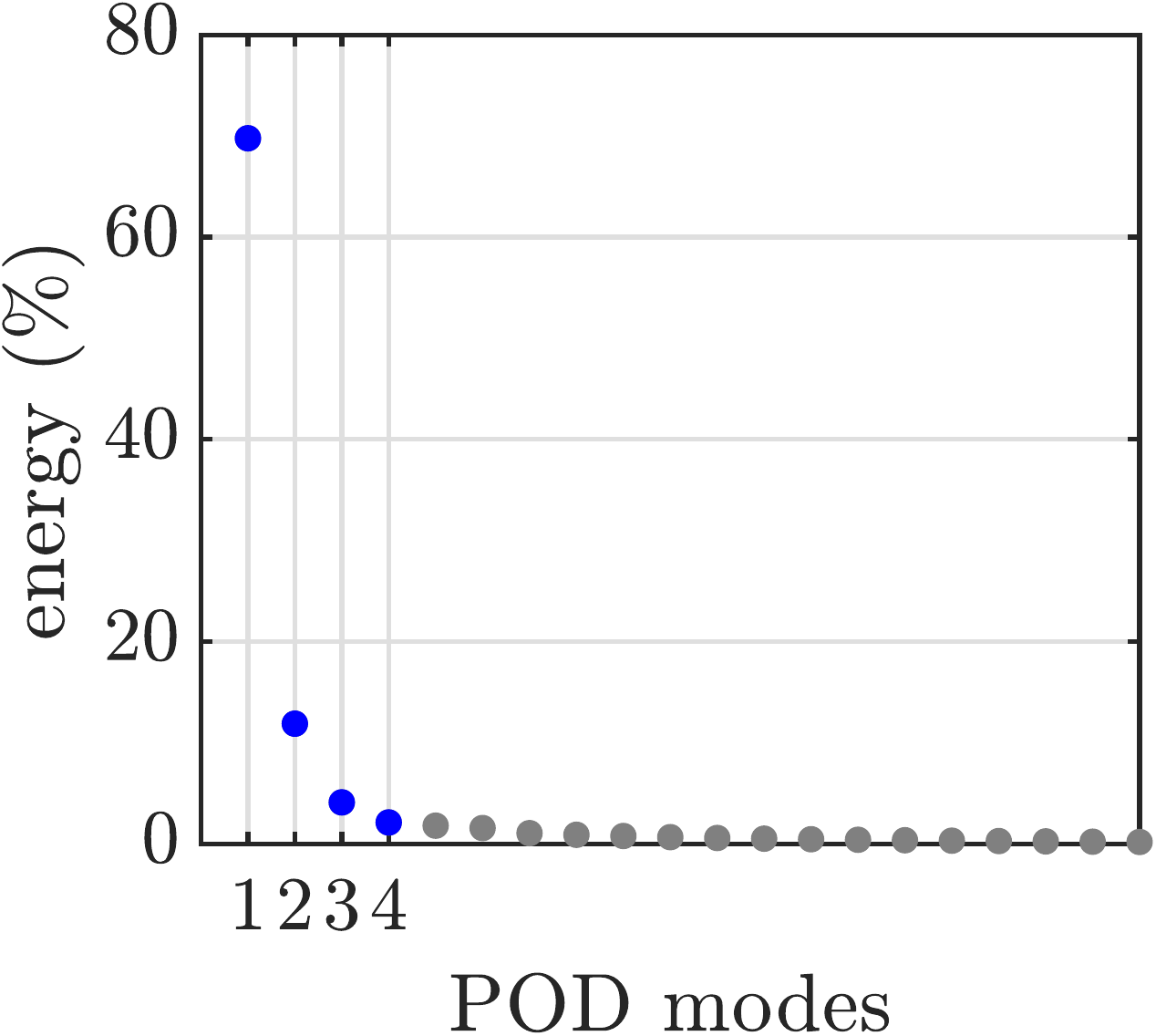}
\includegraphics[width=0.32\linewidth]{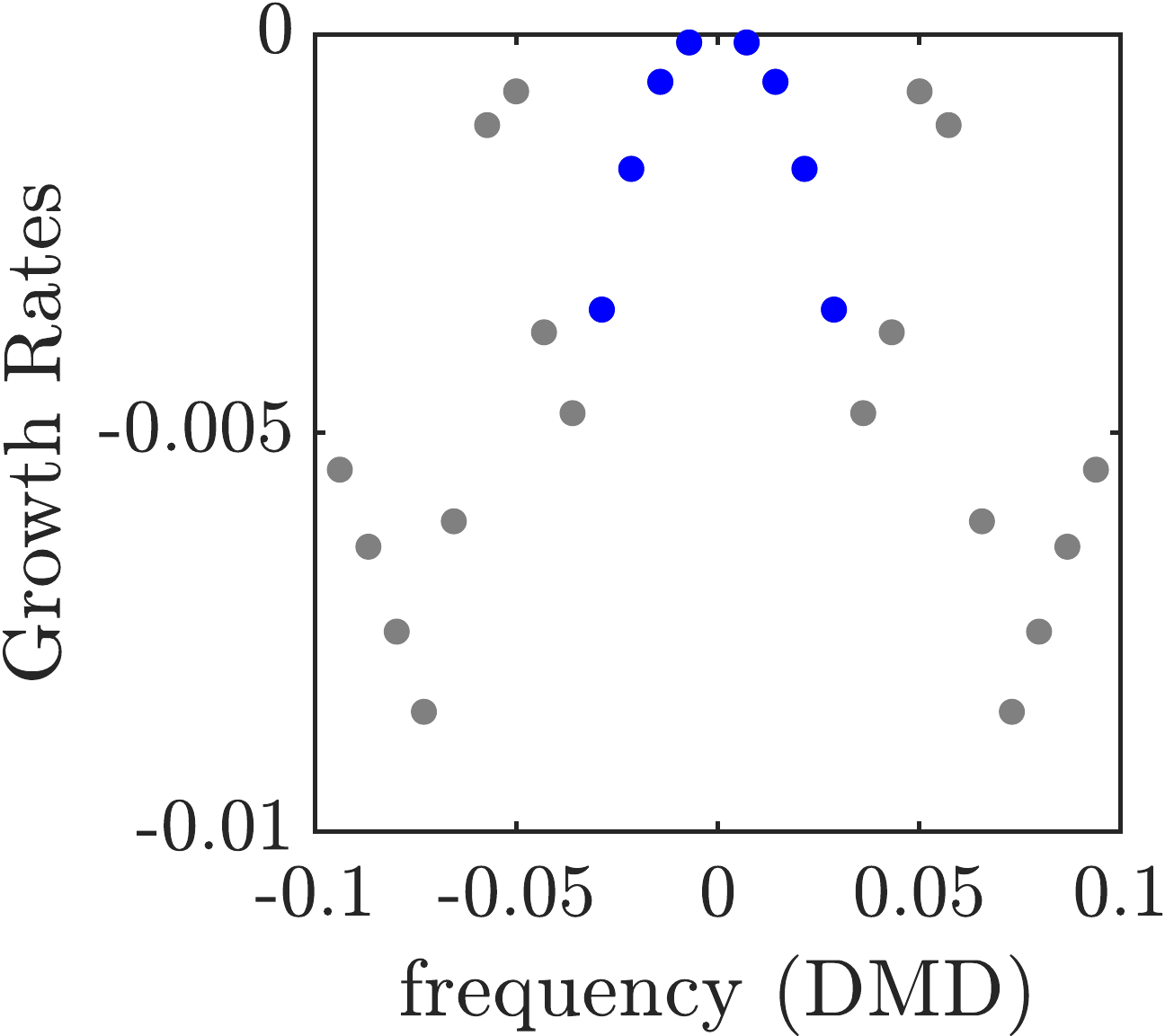}
\includegraphics[width=0.32\linewidth]{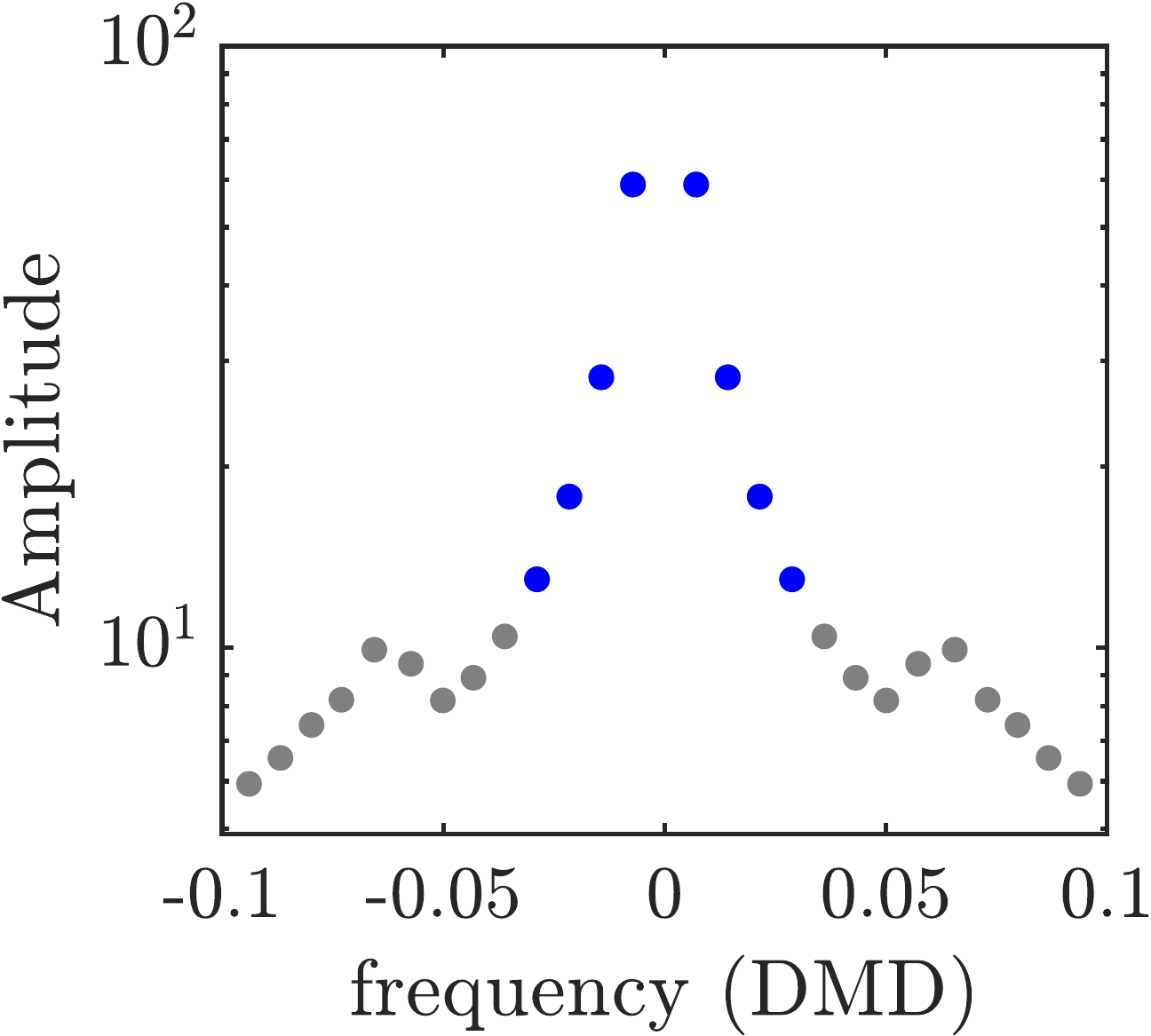}
\put(-377,102){({\it d\hspace{1pt}})}
\put(-249,102){({\it e\hspace{1pt}})}
\put(-124,102){({\it f\hspace{1pt}})}
\end{center}
\caption{\RR A comparison of POD, DMD and EWT modes. The POD (a) and DMD modes (b) are extracted based on the velocity data $u(x,y,z,t)$ with the time series starting from the initiation of FST and ending at the snapshot shown in figure~\ref{F2_bypass1}. The EWT modes (c) are solely extracted from the instantaneous $u(x,y,z)$ corresponding to figure~\ref{F2_bypass1}. The isosurfaces are coloured with wall-normal coordinate and defined by $u=-4\times10^{-4}$ of the extracted POD/DMD modes. 
The same isosurface levels as in figure~\ref{F_bypass5} are used for EWT modes. Panel (d) provides the energy distribution among POD modes, (e) and (f) give the growth rate and amplitude of the DMD modes. The blue dots mark four (pairs of) dominating POD/DMD modes shown in panel (a) and (b).}
\label{F_POD}
\end{figure}
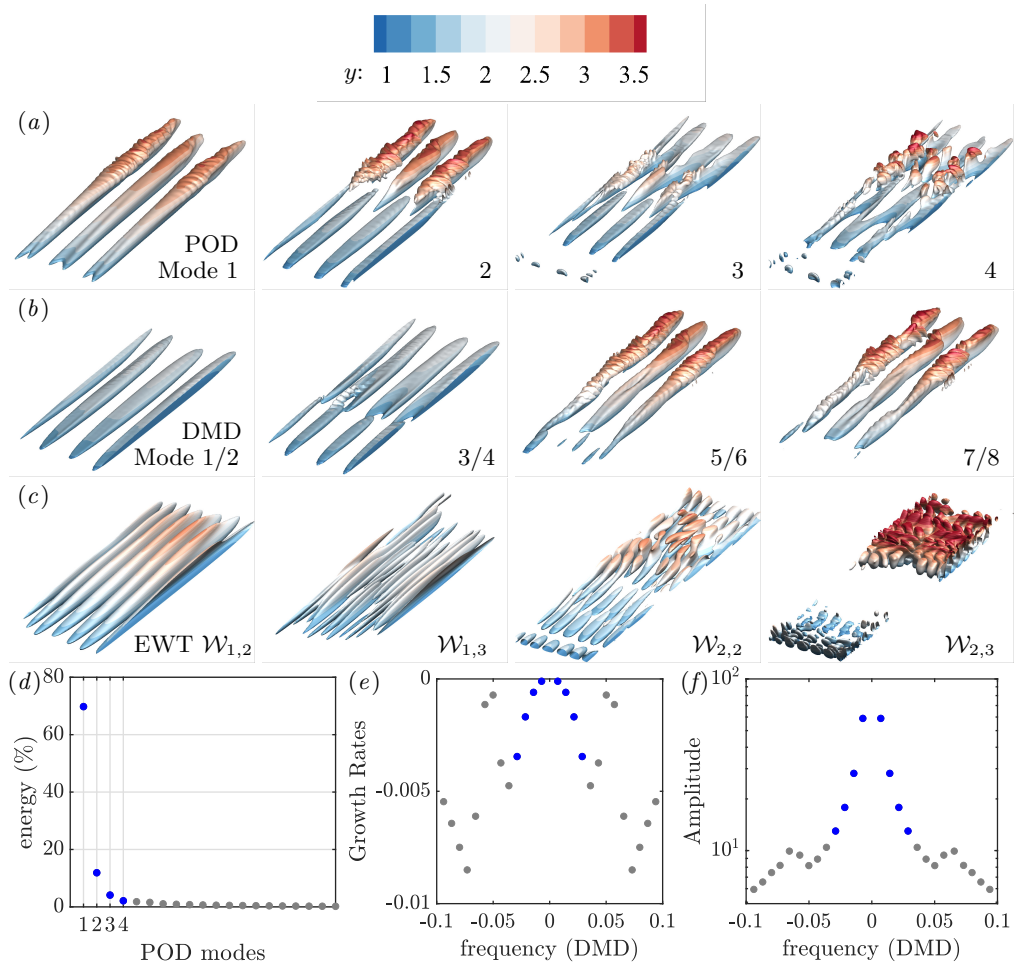

{\RR The flow case corresponds to the example shown in \S~\ref{S32}. Time series as inputs for POD and DMD start from the initialisation of FST, encompassing the development of streamwise-elongated streaks, their secondary instabilities, and end at the snapshot shown in figure~\ref{F_bypass5} where hairpin vortices appear.  Figure~\ref{F_POD} shows the decomposed flows using POD, DMD and EWT, respectively. These modes are illustrated with isosurfaces of the streamwise velocity. From Figure~\ref{F_POD}(d) it is evident that the first four POD modes cover most of the energy as displayed in panel (a). As can be seen, the most energetic structure is streamwise-elongated streaks. Mode 2 \& 3 displayed a mixed structure of streaks and their instabilities. Mode 4, on the other hand, extracts smaller scales associated with the hairpin vortices. The DMD modes appear in pairs and are sorted by their frequencies. As expected, lower frequencies  (Mode 1 \& 2 and 3 \& 4) stand for streaks while the instabilities are captured in higher modes. The growth rates and amplitudes of DMD modes shown in panel (e) and (f) provide a soft criterion as a measure of the modes' significance.}

{\RR The EWT modes are obtained solely from a single snapshot corresponding to figure~\ref{F2_bypass1} and \ref{F_bypass5}. Without resorting to time series, the flows are decomposed according to its spatial scales. As a result, the streaks, their secondary instabilities and the bypass nature of smaller scales are distinctly  identified.} 

{\RR On the aspect of outputs, all three methods can lead to an arbitrary number of modes. A notable question arises as what is the effectiveness of the modes to characterise the flow. For example, how many modes would be sufficient to represent the flow and what is the criteria for selecting dominant modes? This has been recognised as a weakness of modal based flow analysis \citep{Taira2017}.  POD modes are arranged according to their energy, which is a meaningful indication of importance. However, highly nonlinear problems may require a large number of POD modes to cover the majority of the energy \citep{murata2020nonlinear}. On the other hand, a single correct way to rank DMD modes is absent though the growth rate and amplitude provide means of measurement. With regard to EWT, there is not a concomitant indicator (\eg energy, frequency, growth rate) associated with the modes. Instead of producing a large number of modes and choosing a few important ones (\eg in POD and DMD), EWT aims at generating a limited number of modes, differing only in its spatial scales. We have shown that for flows with compact Fourier supports, $M_{x}=M_{z}=3$ gives satisfactory results. However, for flows with a broader spatial spectrum, one may need to increase the number of modes to reduce the mixing of scales in extracted modes. 
}

\begin{figure}
\begin{center}
\includegraphics[width=0.32\linewidth]{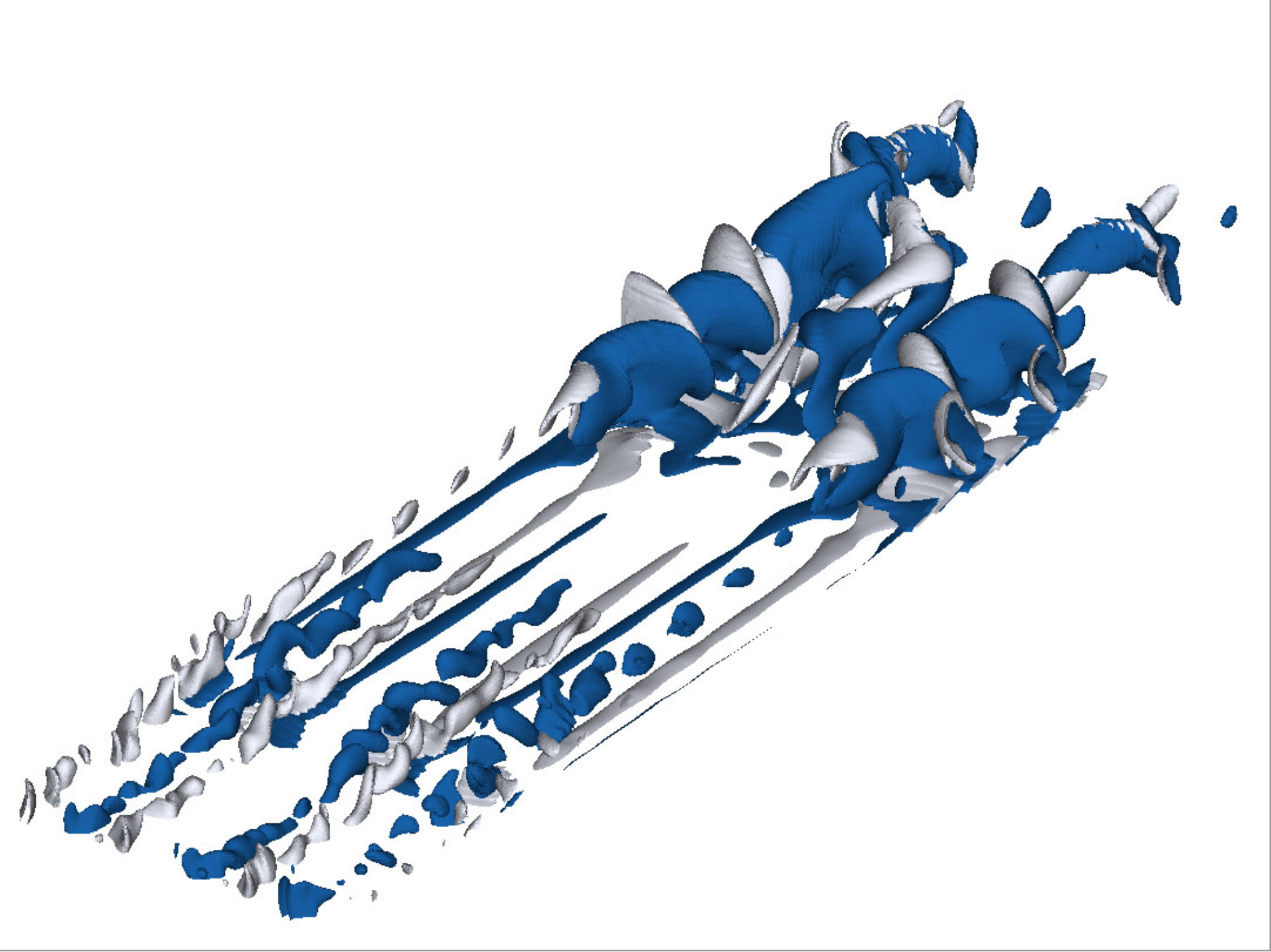}
\includegraphics[width=0.32\linewidth]{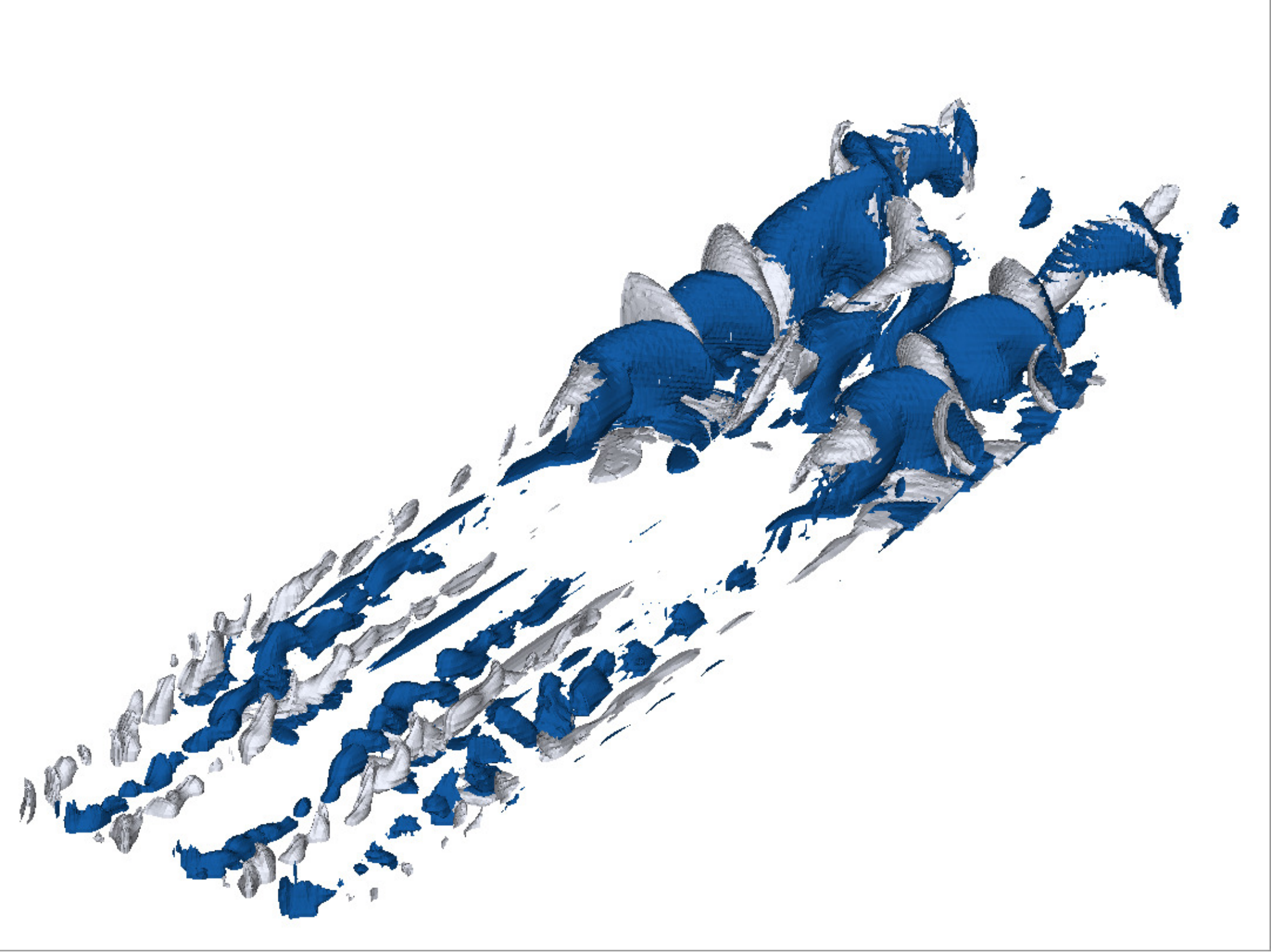} 
\includegraphics[width=0.32\linewidth]{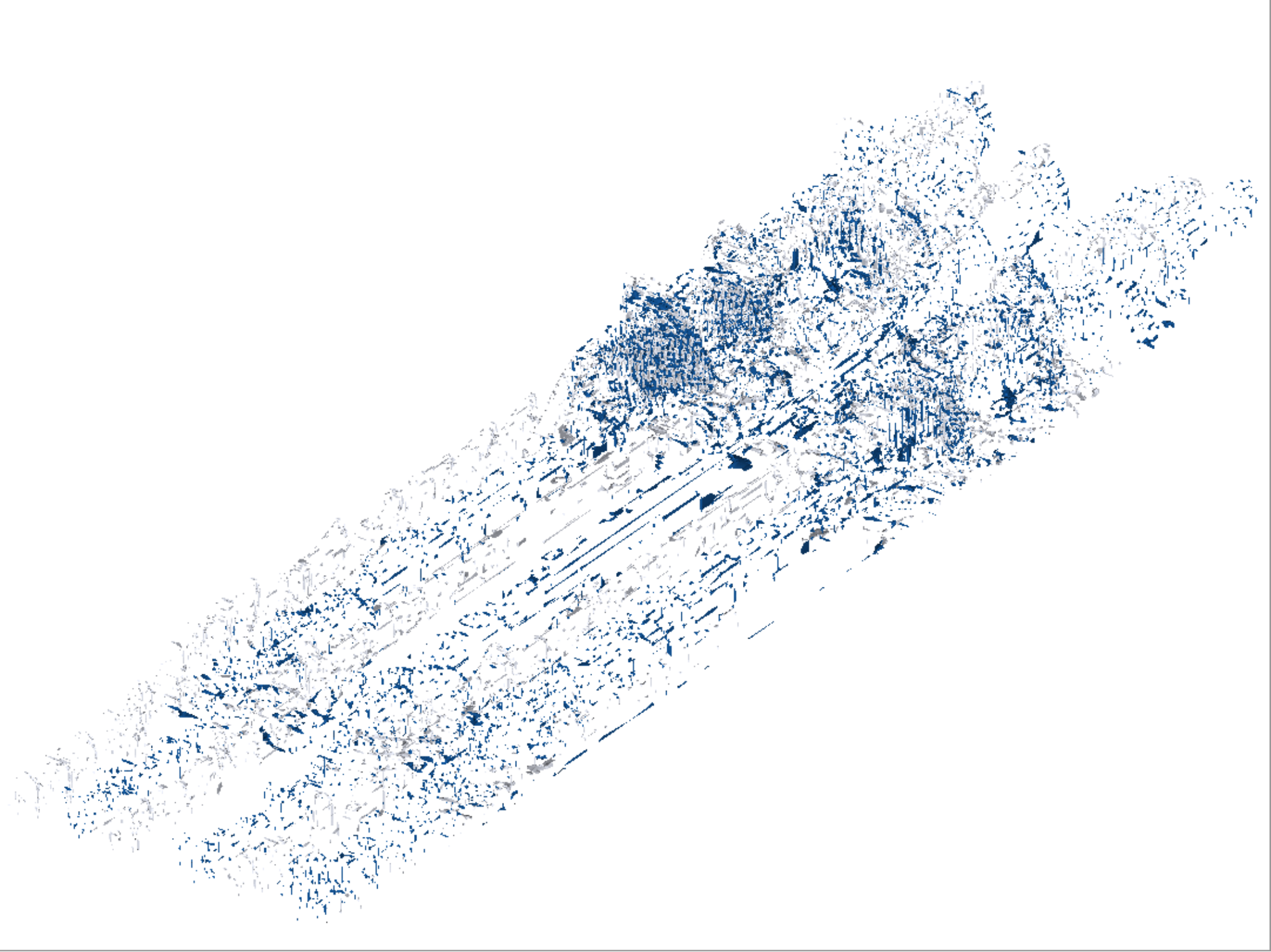}
\put(-300,8){input: $\omega_{x}$}
\put(-176,12){coherent}
\put(-181,4){component}
\put(-56,12){incoherent}
\put(-57,4){component}
\end{center}
\caption{\RR Coherent vortex extraction (CVE) of the transitional boundary layer. The input is the streamwise vorticity $\omega_{x}$. Isosurfaces are defined and coloured by $\omega_{x}=\pm0.1$ for the input and coherent component, while $\omega_{x}=\pm0.06$ is used for the incoherent counterpart to increase visibility.}
\label{CVE}
\end{figure}

{\RR EWT shares a similar theoretical basis, \ie~the multi-resolution analysis, with CVE \citep{farge2001coherent}. In CVE, wavelet coefficients are first obtained for the vorticity field using orthogonal wavelet transform. These coefficients are further divided into two groups (coherent and incoherent) according to their amplitudes. The coherent and incoherent flow components are therefore reconstructed through the inverse wavelet transform. By construction, the objective of CVE is to extract coherent structures based on the wavelet denoising method rather than decomposing a flow into several modes. Results of CVE for the boundary layer transitional flow are provided in figure~\ref{CVE}. For this case, the coherent structure is captured by only 0.8\% of the total wavelet coefficients, reflecting its sparsity distribution in the wavelet space. }

{\RR As such, it becomes clear that EWT provides a new decomposition strategy, as shown in table~\ref{table1}. Furthermore, EWT does not depend on a separation-of-variable strategy, which leads to its advantage dealing with non-stationary flows or traveling wave problems.  }

\begin{table}
\begin{center}\def~{\hphantom{0}}
\begin{tabular}{ccc}
      Methods      & Input  &  Decomposition strategy  \\[3pt]
      POD   		& flow data as time series & by energy \\
      DMD           & flow data as time series & by frequency and growth rate \\
      CVE            & instantaneous flow data & by amplitude of wavelet coefficients \\ 
      EWT           & instantaneous flow data or its visualisation & by averaged spatial Fourier supports 
\end{tabular}
\caption{{\RR A comparison of POD, DMD, CVE and EWT.}}
\label{table1}
\end{center}
\end{table}

\section{Concluding remarks}\label{S4}
In this paper, we proposed an image-based flow decomposition using the 2D tensor empirical wavelet transform (EWT)\citep{Gilles2013,Gilles2014}, which was initially devised for image processing. According to the Fourier supports of the input data, a set of adaptive filter banks (an orthogonal frame) is built to perform the decomposition. A first example considers the interactions between a 2D wake and a jet plume, where only experimental flow visualisations are available. The EWT modes correctly isolate the jet and wake components and their instabilities. In the spirit of 2D tensor EWT, the visualisation's principle direction shall coincide with one of the decomposition directions. We show that this direction can be accurately determined using the shadow mode of EWT ($\mathcal{W}_{1,1}$). The second example considers an early-stage boundary layer transition subject to FST, where direct numerical simulation provided full data-set. For both inputs of 3D instantaneous flow data and its visualisation, the EWT modes distinctly extract streamwise-elongated streaks, their secondary instabilities and smaller scales. A comparison with biGlobal stability analysis justifies the EWT modes that characterise the secondary instabilities. The bypass nature of smaller scales is also captured by EWT modes based on the 3D instantaneous flow data.


Compared to the prevailing data-based methods on flow decomposition (POD, DMD, CVE, to name a few), EWT offers a new strategy to adaptively decompose a flow from its averaged Fourier supports. An instantaneous flow or its visualisation is thus readily decomposed without resorting to its time series. The method also functions well on non-stationary flows or traveling wave problems by extracting fluid physics that are localised from the input. Still, it would be less effective to flows with broader Fourier supports, \eg fully developed turbulent flows. The number of EWT modes need to be in line with the spatial spectrum of the flow. Future development of the method can be extended to account for temporal spectrum and the prediction of flow evolution.

\section*{Declaration of interests}
\noindent The authors report no conflict of interests.

\section*{Acknowledgement}
This investigation was funded by the European Union’s Horizon 2020  \includegraphics[width=0.03\linewidth]{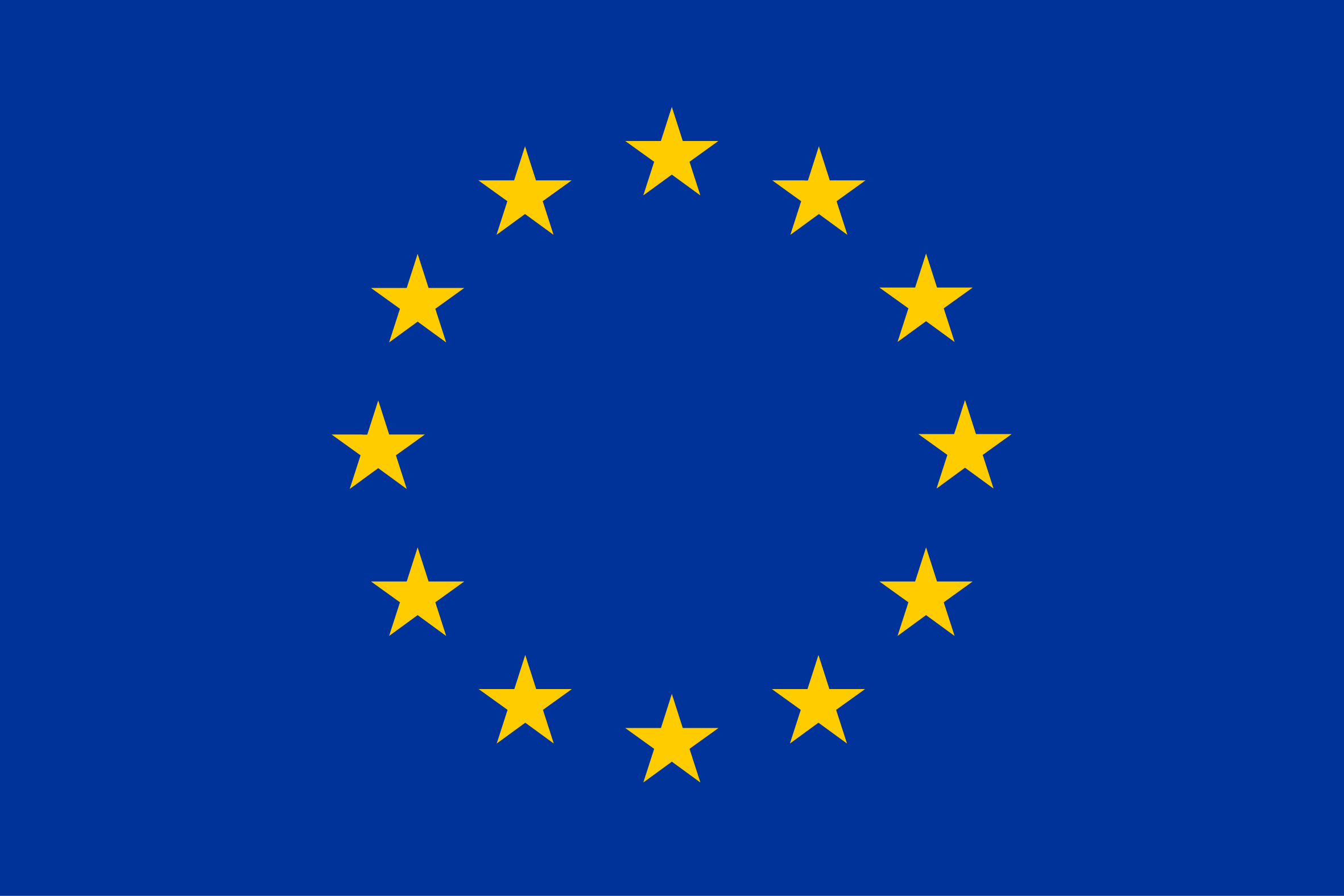} future and emerging technologies programme with agreement No. 828799.

\appendix\section{Scaling function and empirical wavelets of EWT}\label{appA}
The scaling function $\hat{\phi}_{m}$ and empirical wavelets $\hat{\psi}_{m}$ of EWT \citep{Gilles2013} are given by
\begin{equation}\label{ewt1}
\hat{\phi}_{m}\left(\omega\right)=\begin{cases}
1 & \mathrm{if}\left|\omega\right|\leq\left(1-\gamma\right)\omega_{m}\\
\cos\left[\frac{\pi}{2}\beta\left(\frac{\left|\omega\right|-\left(1-\gamma\right)\omega_{m}}{2\gamma\omega_{m}}\right)\right] & \mathrm{if}\left(1-\gamma\right)\omega_{m}\leq\left|\omega\right|\leq\left(1+\gamma\right)\omega_{m}\\
0 & \mathrm{otherwise},
\end{cases}
\end{equation}

\begin{equation}\label{ewt2}
\hat{\psi}_{m}\left(\omega\right)=\begin{cases}
\sin\left[\frac{\pi}{2}\beta\left(\frac{\left|\omega\right|-\left(1-\gamma\right)\omega_{m}}{2\gamma\omega_{m}}\right)\right] & \mathrm{if}\left(1-\gamma\right)\omega_{m}\leq\left|\omega\right|\leq\left(1+\gamma\right)\omega_{m}\\
1 & \mathrm{if}\left(1+\gamma\right)\omega_{m}\leq\left|\omega\right|\leq\left(1-\gamma\right)\omega_{m+1}\\
\cos\left[\frac{\pi}{2}\beta\left(\frac{\left|\omega\right|-\left(1-\gamma\right)\omega_{m+1}}{2\gamma\omega_{m+1}}\right)\right] & \mathrm{if}\left(1-\gamma\right)\omega_{m+1}\leq\left|\omega\right|\leq\left(1+\gamma\right)\omega_{m+1}\\
0 & \mathrm{otherwise},
\end{cases}
\end{equation}
\begin{equation}
{\rm with~} \beta\left(x\right)=\begin{cases}
0 & \mathrm{if}\:x\leq0\\
x^{4}\left(35-84x+70x^{2}-20x^{3}\right) & \mathrm{if}\:0<x<1\\
1 & \mathrm{if}\:x\geq1.
\end{cases}
\end{equation}
\begin{equation}
\mathrm{if\:}\gamma<\underset{m}{\min}\left(\frac{\omega_{m+1}-\omega_{m}}{\omega_{m+1}+\omega_{m}}\right),\mathrm{then}\stackrel[k=-\infty]{\infty}{\sum}\left(\left|\hat{\phi}_{1}\left(\omega+2k\pi\right)\right|^{2}+\stackrel[m=1]{N}{\sum}\left|\hat{\psi}_{m}\left(\omega+2k\pi\right)\right|^{2}\right)=1.
\end{equation}
The set $\{ \phi_{1},\left\{ \psi_{m}\right\} _{m=1}^{M-1}\}$ forms a tight and orthogonal frame of $L^{2}\left(\mathbb{R}\right)$. {\RR Note that in the theory of wavelet frame, a frame is termed {\it tight} if the energy of the extracted wavelet coefficients is directly proportional to the original signal with a factor of $A$. Furthermore, if $A=1$, the frame is defined as {\it orthogonal}.}

\bibliographystyle{jfm}
\bibliography{refs}
\end{document}